\documentclass[%
 reprint,
 floatfix,
 amsmath,amssymb,
 aps,
prd,
lengthcheck
]{revtex4-2}

\usepackage{graphicx}
\usepackage{dcolumn}
\usepackage{bm}
\usepackage{hyperref}
\usepackage[mathlines]{lineno}
\usepackage[dvipsnames]{xcolor}
\hypersetup{
    colorlinks=true,
    linkcolor=blue,
    citecolor = blue,
    filecolor=magenta,      
    urlcolor=cyan,
}

\newcommand{\ecc}{\ensuremath{e}}

\newcommand{\msun}{\ensuremath{M_{\odot}}}

\begin{document}

\title{Eccentric Binary Neutron Star Search Prospects for Cosmic Explorer}

\author{Amber K. Lenon}
\affiliation{Department of Physics, Syracuse University, Syracuse NY 13244, USA}

\author{Alexander H. Nitz}
\affiliation{Max-Planck-Institut f{\"u}r Gravitationsphysik (Albert-Einstein-Institut), D-30167 Hannover, Germany}
\affiliation{Leibniz Universit{\"a}t Hannover, D-30167 Hannover, Germany}%

\author{Duncan A. Brown}
\affiliation{Department of Physics, Syracuse University, Syracuse NY 13244, USA}

\date{\today}

\begin{abstract}
We determine the ability of Cosmic Explorer, a proposed third-generation gravitational-wave observatory, to detect eccentric binary neutron stars and to measure their eccentricity. We find that for a matched-filter search, template banks constructed using binaries in quasi-circular orbits are effectual for eccentric neutron star binaries with $e_{7} \leq 0.004$ ($e_{7} \leq 0.003$) for CE1 (CE2), where $e_7$ is the binary's eccentricity at a gravitational-wave frequency of 7~Hz. We show that stochastic template placement can be used to construct a matched-filter search for binaries with larger eccentricities and construct an effectual template bank for binaries with $e_{7} \leq 0.05$. We show that the computational cost of both the search for binaries in quasi-circular orbits and eccentric orbits is not significantly larger for Cosmic Explorer than for Advanced LIGO and is accessible with present-day computational resources. We investigate Cosmic Explorer's ability to distinguish between circular and eccentric binaries. We estimate that for a binary with a signal-to-noise ratio of 20 (800), Cosmic Explorer can distinguish between a circular binary and a binary with eccentricity $e_7 \gtrsim 10^{-2}$ ($10^{-3}$) at 90\% confidence.
\end{abstract}

\maketitle

\section{\label{sec:Intro}Introduction}

Cosmic Explorer is a proposed third-generation gravitational-wave observatory that will have an order of magnitude improved sensitivity beyond that of Advanced LIGO and will be able to explore gravitational-wave frequencies below 10~Hz~\cite{Reitze:2019iox}. Cosmic Explorer will be able to detect binary neutron stars with a signal-to-noise ratio of $\ge 8$ out to a distance of $\sim 2$~Gpc~\cite{Chen:2017wpg}. Although most of the detected neutron-star binaries will be in circular orbits, measurement of eccentricity in neutron-star mergers allows us to explore their formation and to distinguish between field binaries, which are expected to be circular by the time they are observed~\cite{Peters:1964}, and binaries formed through other channels~\cite{Smarr1976,Canal:1990dz,PortegiesZwart1:1997zn,Postnov:2006hka,Kalogera:2006uj,Kowalska:2010qg,Beniamini:2015uta,Tauris:2017omb,Palmese:2017yhz,Belczynski:2018ptv,Vigna-Gomez:2018dza,Giacobbo:2018etu,Mapelli:2018wys,Andrews:2019vou}.

Dynamical interactions can form binary neutron stars with eccentricity that is measurable, although the predicted rate of these mergers detectable by current gravitational-wave observatories is small~\cite{Lee:2009ca,Ye:2019xvf}. The two binary neutron star mergers observed by Advanced LIGO and Virgo \cite{TheLIGOScientific:2017qsa,Abbott:2020uma} were both detected with searches that use circular waveform templates \cite{Allen:2004gu,Allen:2005fk,Canton:2014ena,Usman:2015kfa,Nitz:2017svb,Sachdev:2019vvd,Cannon:2020qnf,Davies:2020tsx, DalCanton:2020vpm}. Constraints have been placed on the eccentricity of these binaries. At a gravitational-wave frequency of 10 Hz, the eccentricity of GW170817 is $e_{10} \leq 0.024$~\cite{Lenon:2020oza}. The eccentricity of GW190425 has been constrained to $e_{10} \leq 0.007$~\cite{Romero-Shaw:2020aaj} by reweighting the results of parameter estimation with circular waveform templates and to $e_{10} \leq 0.048$ using full parameter estimation with eccentric waveform templates~\cite{Lenon:2020oza} (90\% confidence). Ref.~\cite{Romero-Shaw:2020aaj} considered unstable case BB mass transfer as a formation scenario for GW190425, but the measured eccentricity limit was insufficient to confirm this hypothesis. A search for eccentric binary neutron stars in the first and second Advanced LIGO and Virgo observing runs did not yield any candidates~\cite{Nitz:2019spj}.

By extrapolating the upper limit on the rate of eccentric binary neutron stars from LIGO--Virgo observations, Ref.~\cite{Nitz:2019spj} estimates that the A+ upgrade~\cite{Aasi:2013wya} of Advanced LIGO will require between half a year of observation and $\sim 775$ years of observation before the detectable rate is comparable with the optimistic~\cite{Lee:2009ca} and pessimistic~\cite{Ye:2019xvf} rate predictions respectively, and an observation is plausible.  However, with its increased sensitivity and bandwidth, Cosmic Explorer would need at most half a year of observations to achieve a detectable rate comparable to even the pessimistic models~\cite{Nitz:2019spj}.

We investigate the ability of Cosmic Explorer to detect eccentric binary neutron stars and to measure their eccentricity. We find that at an eccentricity $e_{7} = 0.004$, a matched-filter search using circular waveform templates begins to lose more than $3\%$ of the signal-to-noise ratio due to mismatch between the circular and eccentric waveforms; this is an order of magnitude smaller than the equivalent limit for Advanced LIGO~\cite{Brown:2009ng,Huerta:2013qb}. We demonstrate that stochastic template placement \cite{Harry:2009ea,Manca:2009xw} can be used to construct a template bank that maintains a fitting factor greater than 97\% to binaries with $e_{7} \le 0.05$. We will use a reference frequency of $7$~Hz in reference to eccentricity unless otherwise stated.

Using template banks constructed for Cosmic Explorer, we estimate the computational cost of matched-filter searches for binary neutron stars in circular and eccentric orbits and find that both are accessible with present-day computational resources. We then estimate the ability of Cosmic Explorer to measure and constrain the eccentricity of detected binary neutron star systems. For a binary neutron star with signal-to-noise ratio 8 (800), Cosmic Explorer will be able to measure eccentricities $\gtrsim 8\times 10^{-3}$ ($8\times 10^{-4}$). 

This paper is organized as follows: In Sec.~\ref{s:circ}, we investigate the ability of a matched-filter search to detect eccentric binary neutron stars in Cosmic Explorer. We calculate the lower-frequency cutoff required to obtain at least $99.9\%$ of the available signal-to-noise ratio for binary neutron stars ($m_{1,2} \in [1,3]\, M_\odot$). Using this frequency cutoff, we use geometric placement to construct a template bank using circular waveforms for Cosmic Explorer that has a fitting factor of $97\%$ and estimate the computational cost of performing a matched-filter search using this bank. In Sec.~\ref{s:ecc} we measure the loss in fitting factor when using a bank of circular waveform with neutron-star binaries with eccentricity $e \le 0.05$. We use stochastic template placement to generate a bank containing circular and eccentric waveforms than has a fitting factor of $96.5\%$ and estimate the computational cost of this eccentric binary search. In Sec.~\ref{s:pe}, we estimate the minimum eccentricity that can be measured by Cosmic Explorer as a function of the signal-to-noise ratio of the detected signal. We compare this to estimates of Advanced LIGO and the eccentricity constraints placed by the detection of GW170817. Finally, in Sec.~\ref{s:conc}, we discuss the implications our results for measurement of eccentric binaries with Cosmic Explorer and extension of our work to higher eccentricities and binary black holes.

\section{\label{s:circ} Binary Neutron Star Searches in Cosmic Explorer}

Cosmic Explorer has a two-stage design~\cite{Reitze:2019dyk, Reitze:2019iox}. The first stage of Cosmic Explorer (CE1) assumes that the detector's core technologies will be similar to those of Advanced LIGO with the sensitivity gain from increasing the detector's arm length from 4~km to 40~km. The second stage (CE2) is a technology upgrade to the CE1 detector that further increases Cosmic Explorer's sensitivity. Estimates of the detector's noise power spectral density $S_n(f)$ are available for both CE1 and CE2~\cite{CE:NoiseCurves}; we consider both detector configurations in our analysis. 

Compared to the low-frequency sensitivity limit of Advanced LIGO, which lies around $10$~Hz, Cosmic Explorer pushes the low-frequency limit of the detector below this limit~\cite{Reitze:2019iox}. As for Advanced LIGO, the detector noise begins to rapidly increase as the gravitational-wave frequency reaches the seismic and Newtonian noise walls at low-frequency. The length of a binary neutron star waveform has a steep power-law dependence on its starting frequency $f_\mathrm{lower}$, with the number of cycles between $f_\mathrm{lower}$ and the coalescence frequency scaling as $f_\mathrm{lower}^{-8/3}$. A binary neutron star waveform that starts at $f_\mathrm{lower} = 3$~Hz has a length of approximately $7$~hours, presenting non-trivial data analysis challenges in searches and parameter estimation. 

\begin{figure}
    \includegraphics[width=1.1\columnwidth]{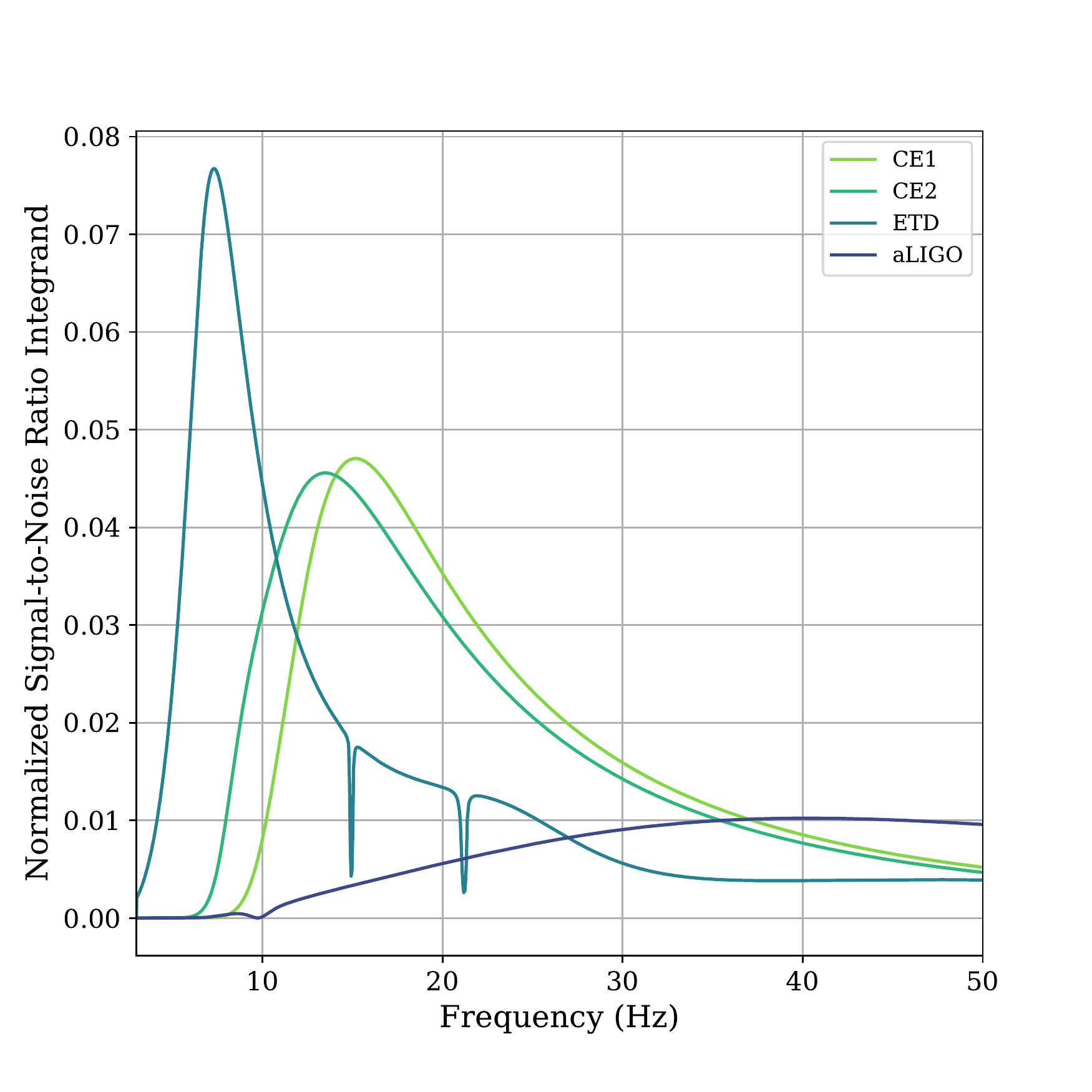}
    \caption{The normalized signal-to-noise ratio integrand as a function of frequency for Cosmic Explorer (CE1/CE2), Einstein Telescope (ETD) and Advanced LIGO (aLIGO). This gives a visual representation of what the matched filter sees when it is integrating up the signal-to-noise ratio. A majority of the signal-to-noise ratio for Cosmic Explorer and Advanced LIGO is accumulated between 10 and 50~Hz, while the signal-to-noise ratio for Einstein Telescope is accumulated below 10~Hz.}
\label{Fig:comp-cost-SNRfrac}
\end{figure}

To determine the optimal starting frequency for binary neutron star searches in Cosmic Explorer, we consider the accumulation of the the signal-to-noise ratio in a matched filter search for a neutron star binary with $m_1 = m_2 = 1.4\,M_\odot$; this accumulates as $f^{-7/3} / S_n(f)$ where $f$ is the gravitational-wave frequency~\cite{Finn:2000hj,Allen:2005fk}. Fig.~\ref{Fig:comp-cost-SNRfrac} shows the normalized signal-to-noise ratio integrand at a given frequency for Cosmic Explorer and Advanced LIGO. Advanced LIGO's most sensitive frequency lies around 40~Hz with almost no detectable signal below 10~Hz. For CE1 and CE2, the peak sensitivity of the detectors to binary neutron stars is shifted to lower frequencies, with a non-trivial amount of signal-to-noise available below 10~Hz.
The fraction of signal-to-noise available drops rapidly as the frequency decreases due to the low-frequency noise wall of Cosmic Explorer.

\begin{figure}
    \includegraphics[width=1.1\columnwidth]{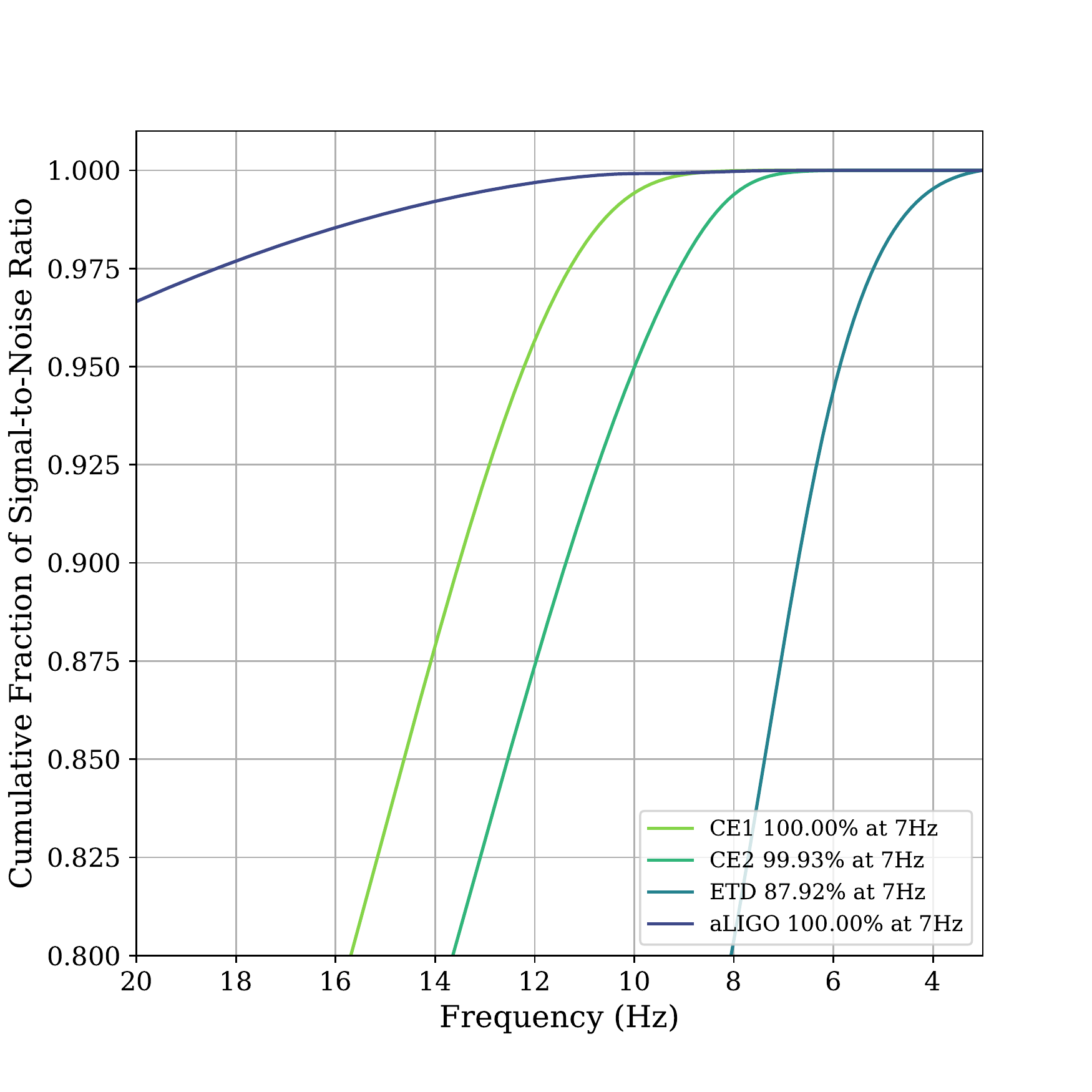}
    \caption{The cumulative fraction of signal-to-noise ratio as a function of frequency. Cosmic Explorer (CE1/CE2) and Advanced LIGO (aLIGO) have accumulated more than 99.9\% of their total signal-to-noise ratio from frequencies above 7~Hz. At 7~Hz, Einstein Telescope (ETD) accumulated more than 85\% of their total signal-to-noise ratio. Since more than 99.9\% of the total signal-to-noise ratio is accumulated, we use a low-frequency cutoff of 7~Hz to generate the waveforms in our template banks.}
\label{Fig:comp-cost-cumulSNR}
\end{figure}

To determine the optimal low-frequency cutoff, we consider the cumulative fraction of the total signal-to-noise ratio as a function of low-frequency cutoff, shown in Fig.~\ref{Fig:comp-cost-cumulSNR}. This is computed by comparing the ratio of the signal-to-noise obtained by integrating from $3$~Hz to a fiducial low-frequency cutoff shown on the ordinate of Fig.~\ref{Fig:comp-cost-cumulSNR}.
We find that for both the CE1 and CE2 detector sensitivity curves, the matched filter accumulates 99.97\% (99.53\%) of the signal-to-noise above 7~Hz for CE1 (CE2). We therefore use 7~Hz as an appropriate low-frequency cutoff for our analysis. At this starting frequency, the length of a binary neutron star waveform is reduced by two orders of magnitude to $4600$~s (77 minutes). For a
waveform of this length, the Doppler frequency modulation due to the diurnal
and orbital motion is $(\Delta f / f ) \sim 10^{-8}$ and can be neglected in
search algorithms. Several search algorithms already exist that can search for waveforms of this length in a computationally efficient manner~\cite{Adams:2015ulm,Sachdev:2019vvd,Cannon:2020qnf}. Similarly, the time dependence of the antenna response due to the Earth's rotation can be neglected as the match between a waveform that neglected the time variation and a waveform that accounted for the variation is 98-99\%. For comparison, we show the same result for the proposed E.U. third-generation detector Einstein Telescope~\cite{Maggiore:2019uih}. We focus on Cosmic Explorer in this work as we have found that existing methodologies are sufficient to effectively address the challenges presented by the increased low-frequency sensitivity of the third generation observatory.  Einstein Telescope has a significantly lower seismic--Newtonian-noise wall than Cosmic Explorer and so searches must be pushed to lower frequencies to accumulate all of the possible signal-to-noise ratio. For an optimal search of Einstein Telescope, this may require addressing how to best account for the time-dependent detector response. 

Using a $7$~Hz low-frequency cutoff, we generate a template bank that can be used to search for binary neutron star mergers with component masses $1.0\, \msun \leq m_1,m_2 \leq 3.0\,\msun$. We first generate a template bank for binaries with zero eccentricity and component spin using the standard hexagonal lattice method of template placement~\cite{Owen:1995tm,Owen:1998dk,Cokelaer:2007kx,Brown:2012qf}. The template bank is constructed so that it has a fitting factor of $97\%$~\cite{Apostolatos:1995pj}. We find that the number of templates required for the CE1 (CE2) sensitivity is $130,000$ ($209,000$) to achieve a fitting factor of $97\%$. A template bank generated using the Advanced LIGO sensitivity and a $10$~Hz low-frequency cutoff contains $77,000$ points. Since the CE1 (CE2) template banks are only a factor of 1.7 (2.8) larger than the equivalent template bank for Advanced LIGO, we do not expect significant computational challenges executing these searches. We certainly expect no obstacles to implementing real-time searches a decade or more from now when Cosmic Explorer will be operational.

Before constructing a template bank for binaries with eccentricity, we determine how effective the non-eccentric template bank is at detecting signals from eccentric binary neutron star sources. We measure the match
\begin{equation}
M = \mathrm{max}_{\phi_\mathrm{c}, t_\mathrm{c}} \frac{(s|h)}{\sqrt{(s|s)}\sqrt{(h|h)}}
\end{equation}
between a random set of eccentric gravitational-wave signals $s$ and the templates $h$, where $(\cdot|\cdot)$ denotes the noise weighted inner product as defined in Ref.~\cite{Allen:2005fk}. The match is maximized over the coalescence time $t_\mathrm{c}$ and coalescence phase $\phi_\mathrm{c}$~\cite{Finn:2000hj}. We maximize the match for each template over the bank to obtain the bank’s fitting factor to a population of eccentric signals~\cite{Apostolatos:1995pj}. The maximum loss in signal-to-noise ratio that the bank will incur due to its discretization is $1-\mathcal{F}$. 

To model eccentric sources, we use the LIGO Algorithm Library implementation \citep{lalsuite} of TaylorF2Ecc, a frequency-domain post-Newtonian model with eccentric corrections. This waveform is accurate to 3.5 pN order in orbital phase \citep{Buonanno:2009zt}, 3.5 pN order in the spin-orbit interactions \citep{Bohe:2013cla}, 2.0 pN order in spin-spin, quadrupole-monopole, and self-interactions of individual spins \citep{Mikoczi:2005dn,Arun:2008kb}, and 3.0 pN order in eccentricity \citep{Moore:2016qxz}. To model non-eccentric waveforms, we use the restricted TaylorF2 approximant, accurate to the same post-Newtonian order. TaylorF2Ecc does not include the merger and ringdown of the signal or depend on the argument of periapsis. Since our study is restricted to binary neutron stars, the merger and ringdown occur at frequencies of order $10^3$~Hz, which is significantly above the frequencies where the majority of the signal-to-noise ratio is accumulated, as shown in Fig.~\ref{Fig:comp-cost-SNRfrac}. Since we restrict our study to relatively low eccentricites, we neglect $\mathcal{O}(e)$ corrections to the waveform amplitude and oscillatory contributions to the waveform phase and hence the argument of periapsis does not enter the waveform computation. Ref.~\cite{Moore:2016qxz} has demonstrated that this does not significantly affect the signal for the cases that we study here.

\begin{figure}
    \includegraphics[width=1.1\columnwidth]{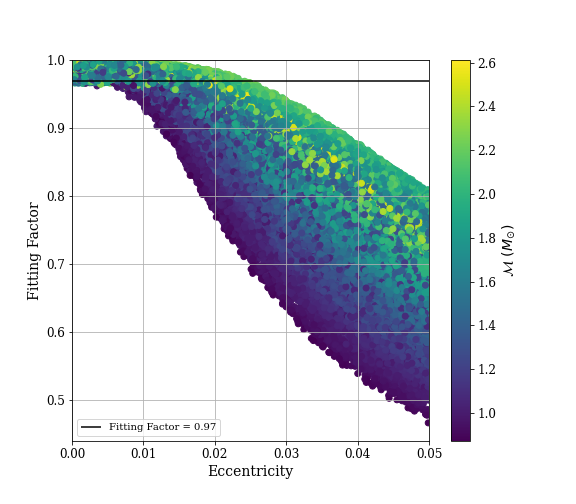}
    \caption{The fitting factor as a function of eccentricity correlated with chirp mass for CE1. A non-eccentric template bank was used to calculate the fitting factor. For Cosmic Explorer the fitting factor decreases for increasing values of eccentricity. The non-eccentric template bank is effective in detecting eccentric systems with a fitting factor greater than 97\% for $\ecc \lesssim 0.004$.}
\label{Fig:ff-circ-CE1}
\end{figure}

\begin{figure}
    \includegraphics[width=1.15\columnwidth]{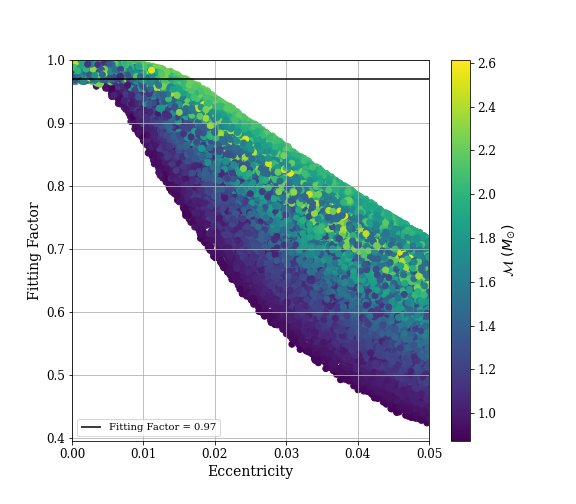}
    \caption{As in Fig.~\ref{Fig:ff-circ-CE1}, but we use the CE2 noise curve. A non-eccentric template bank was used to calculate the fitting factor. The non-eccentric template bank is effective in detecting eccentric systems with a fitting factor greater than 97\% for $\ecc \lesssim 0.003$.}
\label{Fig:ff-circ-CE2}
\end{figure}

We test the template bank against $120,000$ simulated signals that have detector-frame component masses $1.0\,\msun \leq m_1,m_2 \leq 3.0\,\msun$ and eccentricity $0 \le e \le 0.05$. The results of the simulation are shown in Fig.~\ref{Fig:ff-circ-CE1} and Fig.~\ref{Fig:ff-circ-CE2} for CE1 and CE2, respectively. If the population of neutron star binaries has eccentricity less than 0.004 (0.003) in CE1 (CE2), then the non-eccentric template banks achieve a fitting factor of $97\%$ and are effectual. However, for sources with larger eccentricities the effectualness of the template bank begins to rapidly decline; the effectualness of a non-eccentric binary neutron star bank fails at an eccentricity an order of magnitude lower than that of Advanced LIGO~\cite{Brown:2009ng, Huerta:2013qb}. To recover these signals, it is necessary to construct a template bank that captures eccentricity. We consider this in the next section.

\section{\label{s:ecc} Extension to Eccentric Template Banks}

\begin{figure}
    \includegraphics[width=1.1\columnwidth]{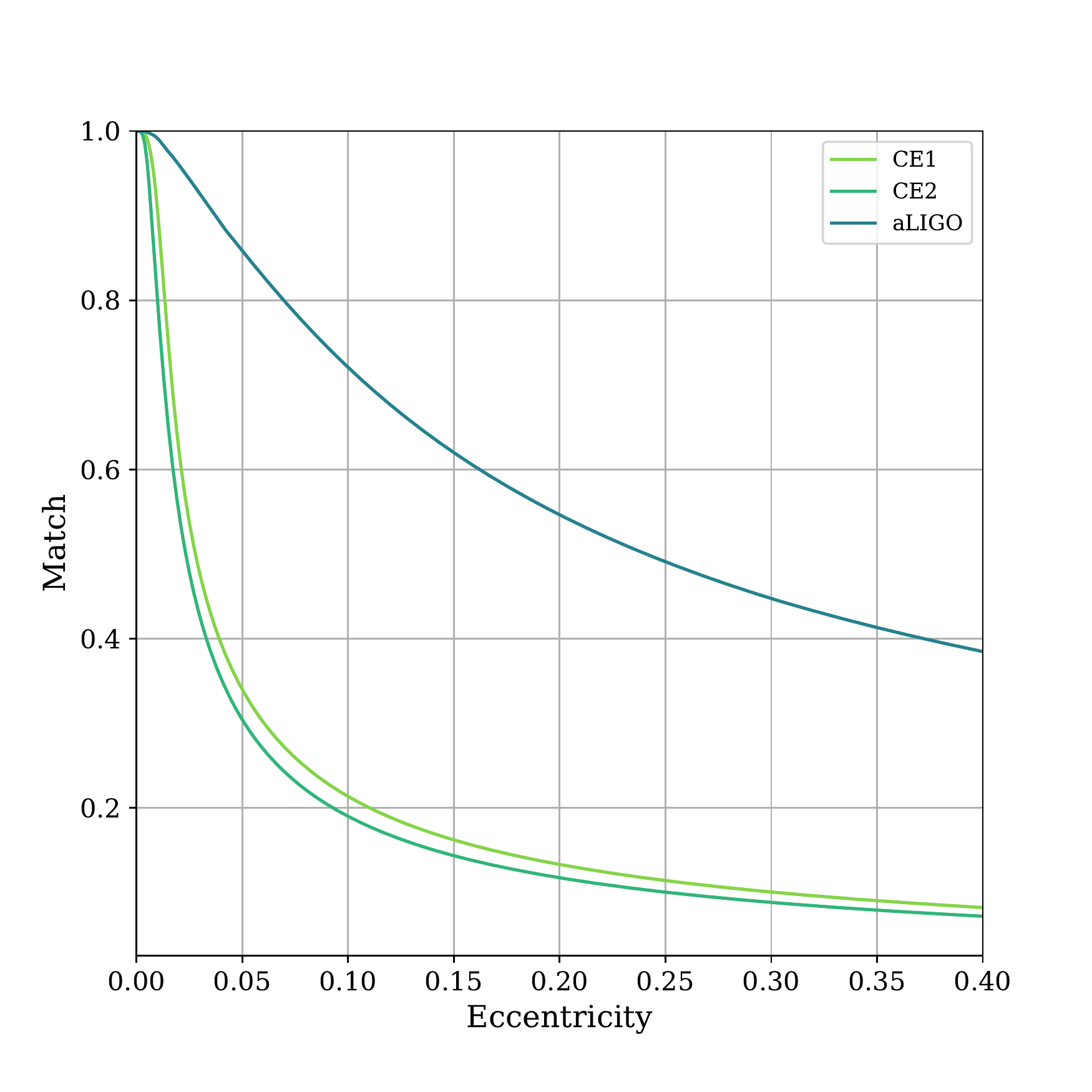}
    \caption{The match as a function of eccentricity for Cosmic Explorer (CE1/CE2) and Advanced LIGO (aLIGO). This gives a representation of the match between a circular waveform and an eccentric waveform for various eccentricities. The match for Cosmic Explorer at an eccentricity of 0.05 is about a factor of 3 smaller than that of Advanced LIGO.}
\label{Fig:eccen-lim}
\end{figure}

The number of templates in an eccentric bank will depend on the bandwidth of the detector and the upper eccentricity boundary of the bank. To visualize the dependency on detector bandwidth, Fig.~\ref{Fig:eccen-lim} shows the eccentricity ambiguity function for a $m_1 = m_2 = 1.4\,\msun$ binary. This shows how quickly the loss in signal-to-noise ratio (match) changes as the eccentricity increases from $0$ to $0.4$ (referenced to 7~Hz). Without the use of eccentric templates, the match for CE1 (CE2) decreases to 34\% (30\%) at $e = 0.05$. In contrast, the Advanced LIGO match decreases much more slowly, reaching 38\% at $e = 0.4$. Consequently, the density of an eccentric template bank will be significantly greater for Cosmic Explorer than Advanced LIGO.

Searches for eccentric binary neutron stars in Advanced LIGO used a template bank that covers the eccentricity range $0 \le e \le 0.4$ (referenced to 10~Hz)~\cite{Nitz:2019spj}; this bank contained $350,000$ templates. To generate template banks of comparable density in eccentricity for Cosmic Explorer, we set the upper eccentricity of the template bank to $e=0.05$ and keep the mass boundaries at $1.0\,\msun \leq m_1,m_2 \leq 3.0\,\msun$ and the lower-frequency cutoff at 7~Hz. We then generated a template bank for eccentric gravitational-wave signals in this region using the stochastic placement method~\cite{Harry:2009ea,Manca:2009xw} with a fitting factor of $96.5\%$. 

\begin{figure}
    \includegraphics[width=1.1\columnwidth]{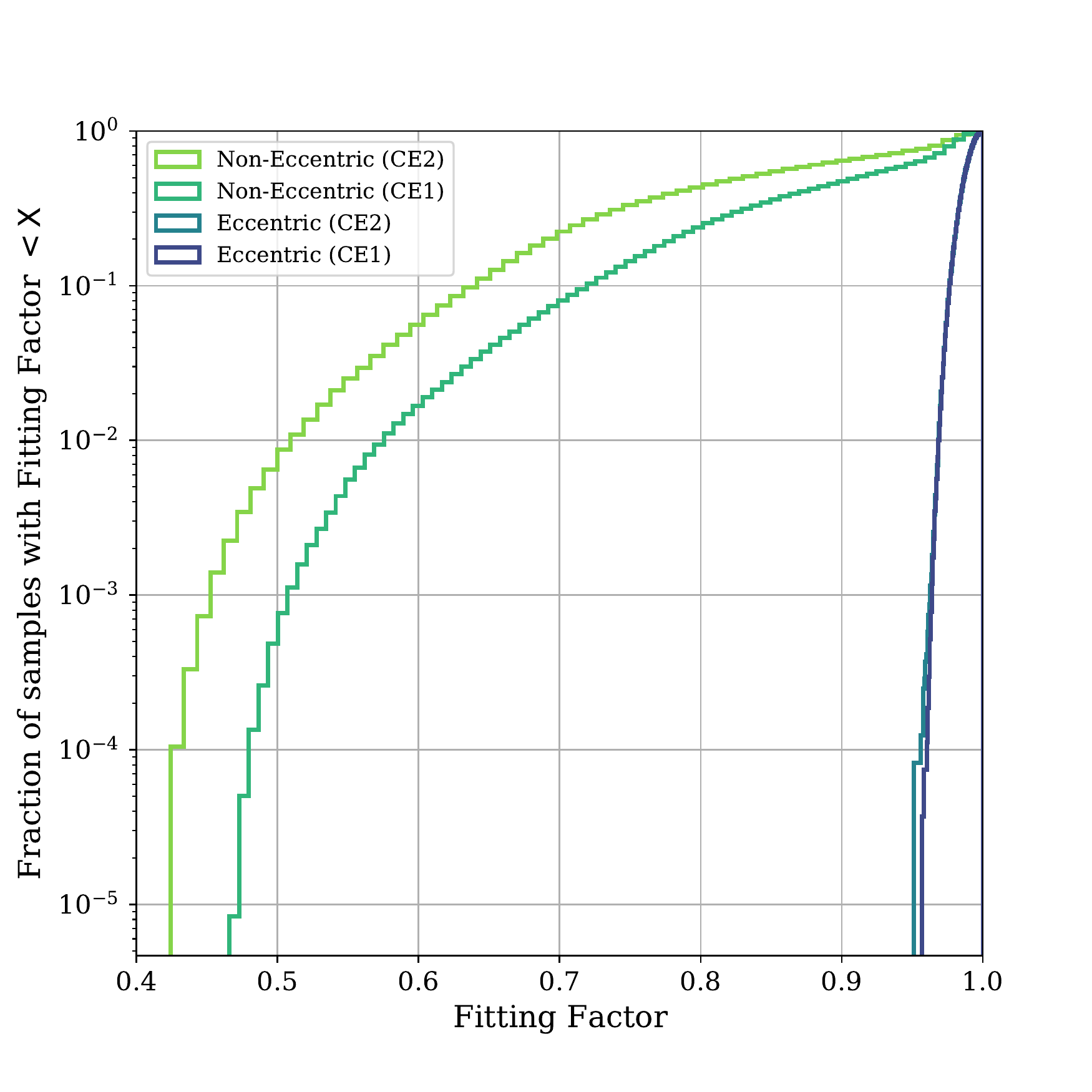}
    \caption{A cumulative histogram that shows the fraction of points where the fitting factor is less than the value on the x-axis for each template bank. Using the eccentric template bank, a majority of the samples are at a fitting factor $\gtrsim 95\%$. For our eccentricity range, the eccentric template banks appear to do a better job at detecting eccentric systems than the non-eccentric template banks.}
\label{Fig:cumul-hist}
\end{figure}

We test the eccentric template bank against $25,000$ simulated signals with detector-frame component masses uniformly distributed between $1.0\,\msun \leq m_1,m_2 \leq 3.0\,\msun$ and eccentricity uniformly distributed between $0 \le e \le 0.05$. The resulting fitting factor of these bank is shown in Fig.~\ref{Fig:cumul-hist}, with the fitting factor of the non-eccentric bank as reference. This result shows that the stochastic bank placement is effectual for signals with eccentricity in the target region, as all signals can be recovered with a fitting factor of $\gtrsim 95\%$ both the CE1 and CE2 banks. The number of eccentric templates generated using the CE1 (CE2) sensitivity is $1,900,000$ ($6,400,000$), an order of magnitude larger than the non-eccentric template banks for CE and an order of magnitude larger than the the Advanced LIGO eccentric bank. We consider the size of a template bank with $e_{max} = 0.1$ to determine the increase in templates as eccentricity increases. A template bank with $0 \le e \le 0.1$ has $4,500,000$ templates using the CE1 sensitivity, this is twice the size of the template bank we consider in this work. We expect that a bank of this size will present no computational challenges when Cosmic Explorer is operational in the 2030s; searches of similar magnitude are already regularly performed~\cite{Nitz:2020bdb, Nitz:2021mzz}.  

\section{\label{s:pe}Binary Neutron Star Parameter Estimation in Cosmic Explorer}

We can use our results to estimate the constraints that Cosmic Explorer will be able to place on the eccentricity of detected binary neutron stars with parameter estimation. We express this as the signal-to-noise ratio required to distinguish between an eccentric and circular binary at 90\% confidence. This can be interpreted as the minimum detectable eccentricity at a given signal-to-noise ratio, or the upper limit that can be placed on the eccentricity of a circular binary detected at a given signal-to-noise ratio.

To estimate the signal-to-noise required to distinguish between a circular binary and a binary with eccentricity $e$ at 90\% confidence, we use the method~\citet{Baird:2012cu}. This method relies on the fact that parameter estimation identifies regions of parameter space where a waveform is most consistent with the data.  Ref.~\cite{Baird:2012cu} uses the fact that high confidence regions in parameter estimation are associated with regions of high match between signal and template to obtain a relationship between the match and signal-to-noise ratio $\rho$, given by
\begin{equation}
    M(h(\theta),h(\langle\theta\rangle)) \geq 1 - \frac{\chi_k^2(1-p)}{2\rho^2}
    \label{eq:baird}
\end{equation}
where $k$ is the dimension of the parameter space of interest, $\chi^2(1 - p)$ is the chi-square value for which there is $1 - p$ probability of obtaining that value or larger. Here, we set $k=4$ corresponding to intrinsic parameter space of an aligned spin binary neutron star merger with eccentricity ($m_1, m_2, \chi_\mathrm{eff}, e$), where $\chi_\mathrm{eff}$ is the effective spin of the binary, and $p = 0.9$ for 90\% confidence.

For Eq.~(\ref{eq:baird}) to provide a reasonable estimate of the signal-to-noise ratio, the match $M$ must be maximized over the parameters of the signal. For eccentric binaries, there is a known degeneracy between the chirp mass $\mathcal{M} = (m_1  m_2)^{3/5} / (m_1 + m_2)^{1/5}$ of the binary and the eccentricity~\cite{Martel:1999tm,Lenon:2020oza}. Full parameter estimation naturally explores the likelihood and this degeneracy. Here, we use our method of eccentric template placement to place a fine grid of templates and brute-force maximize the match over this template bank to account for the chirp mass--eccentricity degeneracy.

Using this method, we estimate Cosmic Explorer's ability to constrain the eccentricity of a $m_1 = m_2 = 1.4\,\msun$ binary as follows: Using a low frequency cutoff of $7$~Hz, we generate a template bank with binary neutron star component masses $1.399\,\msun \leq m_1,m_2 \leq 1.401\,\msun$, eccentricity $0 \leq e \leq 0.05$, an upper-frequency cutoff of $4096$~Hz, and a minimal match of 99.9999\%. We measure the match between a simulated eccentric gravitational-wave signal with component masses $m_1 = m_2 = 1.4\,\msun$ and eccentricity $0 \leq e \leq 0.05$ and maximize over the chirp mass in the template bank to get the signal-to-noise ratio. From this we determine the signal-to-noise ratio needed to reach a 90\% confidence interval~\cite{Baird:2012cu} to measure the eccentricity.

\begin{figure}
    \includegraphics[width=1.1\columnwidth]{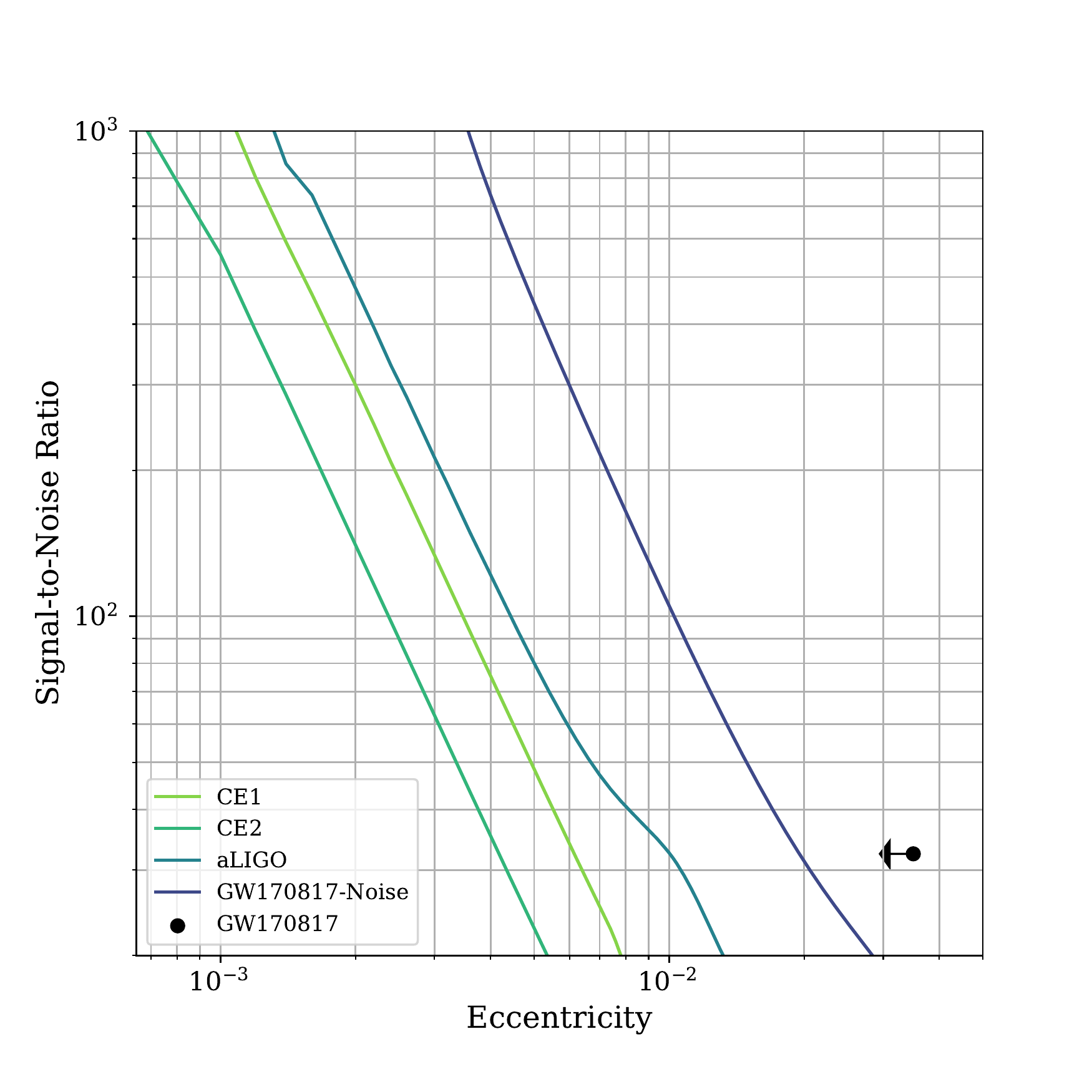}
    \caption{The signal-to-noise ratio as a function of eccentricity. The black dot is at a signal-to-noise ratio of $32.4$~\cite{TheLIGOScientific:2017qsa} and eccentricity of 0.035 at 90\% confidence~\cite{Lenon:2020oza}. For each detector, we show the signal-to-noise ratio needed to resolve the signal with eccentricity on the eccentricity axis at 90\% confidence. For $e \geq 5\times 10^{-3}$, a signal-to-noise ratio of 20 would be needed to resolve the signal at 90\% confidence in Cosmic Explorer. As the eccentricity decreases, the signal-to-noise ratio needed to resolve the signal increases.}
\label{Fig:SNRvEccen}
\end{figure}

We apply the above method using the CE1, CE2, and Advanced LIGO design noise curves to obtain the signal-to-noise ratio as a function of eccentricity required to distinguish between a circular and eccentric binary. To check the accuracy of our estimation, we also compute this function using the detector noise around the time of GW170817 and compare the~\citet{Baird:2012cu} estimate to the 90\% upper limit on the eccentricity of GW170817 computed using full parameter estimation~\cite{Lenon:2020oza}. These result are shown in Fig.~\ref{Fig:SNRvEccen}. First, we note that our method provides a reasonable approximation when comparing to GW170817 and as the eccentricity increases the signal-to-noise ratio needed to resolve the signal decreases, as expected.  Our results suggest that for $e \geq 5\times10^{-3}$ ($8\times 10^{-4}$), a minimum signal-to-noise ratio of 20 (800) would be needed to resolve the signal at 90\% confidence in CE1 and CE2. This is an order of magnitude better than expected from Advanced LIGO operating at design sensitivity.

\section{\label{s:conc}Conclusion}

Our analysis used circular and eccentric template banks to determine the ability of Cosmic Explorer to detect eccentric binary neutron stars and to measure their eccentricity. The circular template banks are effective for detecting eccentric binaries with $e \leq 0.004$ ($e \leq 0.003$) in CE1 (CE2) at a reference frequency of $7$~Hz. However, for larger eccentricities a template bank containing circular and eccentric waveform templates is required. This estimate is an order of magnitude smaller than estimates for Advanced LIGO~\cite{Brown:2009ng, Huerta:2013qb}. We determine the signal-to-noise ratio needed to constrain the eccentricity of a detected neutron star binary signal with 90\% confidence. We also estimate that in Cosmic Explorer to measure binary neutron stars with eccentricity $\gtrsim 5\times10^{-3}$ ($8\times 10^{-4}$) a signal-to-noise ratio of 20 (800) is needed to resolve the signal at a reference frequency of $7$~Hz (90\% confidence). Our method of estimation relies on the calculation of~\citet{Baird:2012cu}; a more accurate determination of this limit requires a full parameter estimation study with accurate eccentric merger waveforms, which is the subject of future work. Accurately constraining the eccentricity of the binary would provide valuable information on the formation of these mergers.

The computational cost of searches with template banks containing higher eccentricities will be challenging in Cosmic Explorer today as the density of the template bank increases with increasing eccentricity (see Fig. 2 of Ref.~\cite{Nitz:2019spj}). However, improvements in technology by the 2030s may make these searches a possibility. Along with the high computational cost, current waveform models for eccentricity break down for $e \geq 0.4$. To accurately search for higher eccentricity neutron-star binaries models that extend to high eccentricities will need to be developed or a burst search will need to be used. In this analysis, we have only considered binary neutron star signals where the measurement of eccentricity is dominated by the inspiral signal. Using an eccentric merger-ringdown waveform~\cite{Huerta:2014eca} an overlap method similar to the method used here, Ref.~\cite{Lower:2018seu}, predict that Cosmic Explorer will be able to distinguish between circular and eccentric waveforms with $e \gtrsim 2 \times 10^{-4}$ for signal-to-noise ratios of order 200. More detailed studies with full parameter estimation and accurate waveforms~\cite{Tanay:2016zog, Huerta:2016rwp, Cao:2017ndf, Huerta:2017kez, Hinder:2017sxy} will be required to further explore these predictions. Understanding the constraints that future observational limits place on eccentric binary formation channels will require computation of the rate as a function of eccentricity from population synthesis. As Cosmic Explorer will be able to aid in the understanding of the physics of binary neutron star mergers it is important to accurately constrain the eccentricity as the number of mergers increases.

\begin{acknowledgments}
We acknowledge the Max Planck Gesellschaft for support and the Atlas cluster computing team at AEI Hannover. DAB thanks National Science Foundation Grants~PHY-1707954 and PHY-2011655 for support. AL thanks National Science Foundation Grant AST-1559694 for support.
\end{acknowledgments}


\begin{thebibliography}{67}%
\makeatletter
\providecommand \@ifxundefined [1]{%
 \@ifx{#1\undefined}
}%
\providecommand \@ifnum [1]{%
 \ifnum #1\expandafter \@firstoftwo
 \else \expandafter \@secondoftwo
 \fi
}%
\providecommand \@ifx [1]{%
 \ifx #1\expandafter \@firstoftwo
 \else \expandafter \@secondoftwo
 \fi
}%
\providecommand \natexlab [1]{#1}%
\providecommand \enquote  [1]{``#1''}%
\providecommand \bibnamefont  [1]{#1}%
\providecommand \bibfnamefont [1]{#1}%
\providecommand \citenamefont [1]{#1}%
\providecommand \href@noop [0]{\@secondoftwo}%
\providecommand \href [0]{\begingroup \@sanitize@url \@href}%
\providecommand \@href[1]{\@@startlink{#1}\@@href}%
\providecommand \@@href[1]{\endgroup#1\@@endlink}%
\providecommand \@sanitize@url [0]{\catcode `\\12\catcode `\$12\catcode
  `\&12\catcode `\#12\catcode `\^12\catcode `\_12\catcode `\%12\relax}%
\providecommand \@@startlink[1]{}%
\providecommand \@@endlink[0]{}%
\providecommand \url  [0]{\begingroup\@sanitize@url \@url }%
\providecommand \@url [1]{\endgroup\@href {#1}{\urlprefix }}%
\providecommand \urlprefix  [0]{URL }%
\providecommand \Eprint [0]{\href }%
\providecommand \doibase [0]{https://doi.org/}%
\providecommand \selectlanguage [0]{\@gobble}%
\providecommand \bibinfo  [0]{\@secondoftwo}%
\providecommand \bibfield  [0]{\@secondoftwo}%
\providecommand \translation [1]{[#1]}%
\providecommand \BibitemOpen [0]{}%
\providecommand \bibitemStop [0]{}%
\providecommand \bibitemNoStop [0]{.\EOS\space}%
\providecommand \EOS [0]{\spacefactor3000\relax}%
\providecommand \BibitemShut  [1]{\csname bibitem#1\endcsname}%
\let\auto@bib@innerbib\@empty
\bibitem [{\citenamefont {Reitze}\ \emph
  {et~al.}(2019{\natexlab{a}})\citenamefont {Reitze} \emph
  {et~al.}}]{Reitze:2019iox}%
  \BibitemOpen
  \bibfield  {author} {\bibinfo {author} {\bibfnamefont {D.}~\bibnamefont
  {Reitze}} \emph {et~al.},\ }\bibfield  {title} {\bibinfo {title} {{Cosmic
  Explorer: The U.S. Contribution to Gravitational-Wave Astronomy beyond
  LIGO}},\ }\href@noop {} {\bibfield  {journal} {\bibinfo  {journal} {Bull. Am.
  Astron. Soc.}\ }\textbf {\bibinfo {volume} {51}},\ \bibinfo {pages} {035}
  (\bibinfo {year} {2019}{\natexlab{a}})},\ \Eprint
  {https://arxiv.org/abs/1907.04833} {arXiv:1907.04833 [astro-ph.IM]}
  \BibitemShut {NoStop}%
\bibitem [{\citenamefont {Chen}\ \emph {et~al.}(2021)\citenamefont {Chen},
  \citenamefont {Holz}, \citenamefont {Miller}, \citenamefont {Evans},
  \citenamefont {Vitale},\  and\ \citenamefont {Creighton}}]{Chen:2017wpg}%
  \BibitemOpen
  \bibfield  {author} {\bibinfo {author} {\bibfnamefont {H.-Y.}\ \bibnamefont
  {Chen}}, \bibinfo {author} {\bibfnamefont {D.~E.}\ \bibnamefont {Holz}},
  \bibinfo {author} {\bibfnamefont {J.}~\bibnamefont {Miller}}, \bibinfo
  {author} {\bibfnamefont {M.}~\bibnamefont {Evans}}, \bibinfo {author}
  {\bibfnamefont {S.}~\bibnamefont {Vitale}},  and\ \bibinfo {author}
  {\bibfnamefont {J.}~\bibnamefont {Creighton}},\ }\bibfield  {title} {\bibinfo
  {title} {{Distance measures in gravitational-wave astrophysics and
  cosmology}},\ }\href {https://doi.org/10.1088/1361-6382/abd594} {\bibfield
  {journal} {\bibinfo  {journal} {Class. Quant. Grav.}\ }\textbf {\bibinfo
  {volume} {38}},\ \bibinfo {pages} {055010} (\bibinfo {year} {2021})},\
  \Eprint {https://arxiv.org/abs/1709.08079} {arXiv:1709.08079 [astro-ph.CO]}
  \BibitemShut {NoStop}%
\bibitem [{\citenamefont {Peters}(1964)}]{Peters:1964}%
  \BibitemOpen
  \bibfield  {author} {\bibinfo {author} {\bibfnamefont {P.~C.}\ \bibnamefont
  {Peters}},\ }\bibfield  {title} {\bibinfo {title} {{Gravitational Radiation
  and the Motion of Two Point Masses}},\ }\href
  {https://doi.org/10.1103/PhysRev.136.B1224} {\bibfield  {journal} {\bibinfo
  {journal} {Phys. Rev.}\ }\textbf {\bibinfo {volume} {136}},\ \bibinfo {pages}
  {B1224} (\bibinfo {year} {1964})}\BibitemShut {NoStop}%
\bibitem [{\citenamefont {{Smarr}}\  and\ \citenamefont
  {{Blandford}}(1976)}]{Smarr1976}%
  \BibitemOpen
  \bibfield  {author} {\bibinfo {author} {\bibfnamefont {L.~L.}\ \bibnamefont
  {{Smarr}}} and\ \bibinfo {author} {\bibfnamefont {R.}~\bibnamefont
  {{Blandford}}},\ }\bibfield  {title} {\bibinfo {title} {{The binary pulsar:
  physical processes, possible companions, and evolutionary histories.}},\
  }\href {https://doi.org/10.1086/154524} {\bibfield  {journal} {\bibinfo
  {journal} {\apj}\ }\textbf {\bibinfo {volume} {207}},\ \bibinfo {pages} {574}
  (\bibinfo {year} {1976})}\BibitemShut {NoStop}%
\bibitem [{\citenamefont {Canal}\ \emph {et~al.}(1990)\citenamefont {Canal},
  \citenamefont {Isern},\  and\ \citenamefont {Labay}}]{Canal:1990dz}%
  \BibitemOpen
  \bibfield  {author} {\bibinfo {author} {\bibfnamefont {R.}~\bibnamefont
  {Canal}}, \bibinfo {author} {\bibfnamefont {J.}~\bibnamefont {Isern}},  and\
  \bibinfo {author} {\bibfnamefont {J.}~\bibnamefont {Labay}},\ }\bibfield
  {title} {\bibinfo {title} {{The origin of neutron stars in binary systems}},\
  }\href {https://doi.org/10.1146/annurev.aa.28.090190.001151} {\bibfield
  {journal} {\bibinfo  {journal} {Ann. Rev. Astron. Astrophys.}\ }\textbf
  {\bibinfo {volume} {28}},\ \bibinfo {pages} {183} (\bibinfo {year}
  {1990})}\BibitemShut {NoStop}%
\bibitem [{\citenamefont {Portegies~Zwart}\  and\ \citenamefont
  {Yungelson}(1998)}]{PortegiesZwart1:1997zn}%
  \BibitemOpen
  \bibfield  {author} {\bibinfo {author} {\bibfnamefont {S.~F.}\ \bibnamefont
  {Portegies~Zwart}} and\ \bibinfo {author} {\bibfnamefont {L.~R.}\
  \bibnamefont {Yungelson}},\ }\bibfield  {title} {\bibinfo {title} {{Formation
  and evolution of binary neutron stars}},\ }\href@noop {} {\bibfield
  {journal} {\bibinfo  {journal} {Astron. Astrophys.}\ }\textbf {\bibinfo
  {volume} {332}},\ \bibinfo {pages} {173} (\bibinfo {year} {1998})},\ \Eprint
  {https://arxiv.org/abs/astro-ph/9710347} {arXiv:astro-ph/9710347}
  \BibitemShut {NoStop}%
\bibitem [{\citenamefont {Postnov}\  and\ \citenamefont
  {Yungelson}(2006)}]{Postnov:2006hka}%
  \BibitemOpen
  \bibfield  {author} {\bibinfo {author} {\bibfnamefont {K.}~\bibnamefont
  {Postnov}} and\ \bibinfo {author} {\bibfnamefont {L.}~\bibnamefont
  {Yungelson}},\ }\bibfield  {title} {\bibinfo {title} {{The Evolution of
  Compact Binary Star Systems}},\ }\href {https://doi.org/10.12942/lrr-2006-6}
  {\bibfield  {journal} {\bibinfo  {journal} {Living Rev. Rel.}\ }\textbf
  {\bibinfo {volume} {9}},\ \bibinfo {pages} {6} (\bibinfo {year} {2006})},\
  \Eprint {https://arxiv.org/abs/astro-ph/0701059} {arXiv:astro-ph/0701059}
  \BibitemShut {NoStop}%
\bibitem [{\citenamefont {Kalogera}\ \emph {et~al.}(2007)\citenamefont
  {Kalogera}, \citenamefont {Belczynski}, \citenamefont {Kim}, \citenamefont
  {O'Shaughnessy},\  and\ \citenamefont {Willems}}]{Kalogera:2006uj}%
  \BibitemOpen
  \bibfield  {author} {\bibinfo {author} {\bibfnamefont {V.}~\bibnamefont
  {Kalogera}}, \bibinfo {author} {\bibfnamefont {K.}~\bibnamefont
  {Belczynski}}, \bibinfo {author} {\bibfnamefont {C.}~\bibnamefont {Kim}},
  \bibinfo {author} {\bibfnamefont {R.~W.}\ \bibnamefont {O'Shaughnessy}},
  and\ \bibinfo {author} {\bibfnamefont {B.}~\bibnamefont {Willems}},\
  }\bibfield  {title} {\bibinfo {title} {{Formation of Double Compact
  Objects}},\ }\href {https://doi.org/10.1016/j.physrep.2007.02.008} {\bibfield
   {journal} {\bibinfo  {journal} {Phys. Rept.}\ }\textbf {\bibinfo {volume}
  {442}},\ \bibinfo {pages} {75} (\bibinfo {year} {2007})},\ \Eprint
  {https://arxiv.org/abs/astro-ph/0612144} {arXiv:astro-ph/0612144}
  \BibitemShut {NoStop}%
\bibitem [{\citenamefont {Kowalska}\ \emph {et~al.}(2011)\citenamefont
  {Kowalska}, \citenamefont {Bulik}, \citenamefont {Belczynski}, \citenamefont
  {Dominik},\  and\ \citenamefont {Gondek-Rosinska}}]{Kowalska:2010qg}%
  \BibitemOpen
  \bibfield  {author} {\bibinfo {author} {\bibfnamefont {I.}~\bibnamefont
  {Kowalska}}, \bibinfo {author} {\bibfnamefont {T.}~\bibnamefont {Bulik}},
  \bibinfo {author} {\bibfnamefont {K.}~\bibnamefont {Belczynski}}, \bibinfo
  {author} {\bibfnamefont {M.}~\bibnamefont {Dominik}},  and\ \bibinfo {author}
  {\bibfnamefont {D.}~\bibnamefont {Gondek-Rosinska}},\ }\bibfield  {title}
  {\bibinfo {title} {{The eccentricity distribution of compact binaries}},\
  }\href {https://doi.org/10.1051/0004-6361/201015777} {\bibfield  {journal}
  {\bibinfo  {journal} {Astron.\ Astrophys.}\ }\textbf {\bibinfo {volume}
  {527}},\ \bibinfo {pages} {A70} (\bibinfo {year} {2011})},\ \Eprint
  {https://arxiv.org/abs/1010.0511} {arXiv:1010.0511 [astro-ph.CO]}
  \BibitemShut {NoStop}%
\bibitem [{\citenamefont {Beniamini}\  and\ \citenamefont
  {Piran}(2016)}]{Beniamini:2015uta}%
  \BibitemOpen
  \bibfield  {author} {\bibinfo {author} {\bibfnamefont {P.}~\bibnamefont
  {Beniamini}} and\ \bibinfo {author} {\bibfnamefont {T.}~\bibnamefont
  {Piran}},\ }\bibfield  {title} {\bibinfo {title} {{Formation of Double
  Neutron Star systems as implied by observations}},\ }\href
  {https://doi.org/10.1093/mnras/stv2903} {\bibfield  {journal} {\bibinfo
  {journal} {Mon. Not. Roy. Astron. Soc.}\ }\textbf {\bibinfo {volume} {456}},\
  \bibinfo {pages} {4089} (\bibinfo {year} {2016})},\ \Eprint
  {https://arxiv.org/abs/1510.03111} {arXiv:1510.03111 [astro-ph.HE]}
  \BibitemShut {NoStop}%
\bibitem [{\citenamefont {Tauris}\ \emph {et~al.}(2017)\citenamefont {Tauris}
  \emph {et~al.}}]{Tauris:2017omb}%
  \BibitemOpen
  \bibfield  {author} {\bibinfo {author} {\bibfnamefont {T.}~\bibnamefont
  {Tauris}} \emph {et~al.},\ }\bibfield  {title} {\bibinfo {title} {{Formation
  of Double Neutron Star Systems}},\ }\href
  {https://doi.org/10.3847/1538-4357/aa7e89} {\bibfield  {journal} {\bibinfo
  {journal} {Astrophys. J.}\ }\textbf {\bibinfo {volume} {846}},\ \bibinfo
  {pages} {170} (\bibinfo {year} {2017})},\ \Eprint
  {https://arxiv.org/abs/1706.09438} {arXiv:1706.09438 [astro-ph.HE]}
  \BibitemShut {NoStop}%
\bibitem [{\citenamefont {Palmese}\ \emph {et~al.}(2017)\citenamefont {Palmese}
  \emph {et~al.}}]{Palmese:2017yhz}%
  \BibitemOpen
  \bibfield  {author} {\bibinfo {author} {\bibfnamefont {A.}~\bibnamefont
  {Palmese}} \emph {et~al.},\ }\bibfield  {title} {\bibinfo {title} {{Evidence
  for Dynamically Driven Formation of the GW170817 Neutron Star Binary in NGC
  4993}},\ }\href {https://doi.org/10.3847/2041-8213/aa9660} {\bibfield
  {journal} {\bibinfo  {journal} {Astrophys. J. Lett.}\ }\textbf {\bibinfo
  {volume} {849}},\ \bibinfo {pages} {L34} (\bibinfo {year} {2017})},\ \Eprint
  {https://arxiv.org/abs/1710.06748} {arXiv:1710.06748 [astro-ph.HE]}
  \BibitemShut {NoStop}%
\bibitem [{\citenamefont {Belczynski}\ \emph {et~al.}(2018)\citenamefont
  {Belczynski} \emph {et~al.}}]{Belczynski:2018ptv}%
  \BibitemOpen
  \bibfield  {author} {\bibinfo {author} {\bibfnamefont {K.}~\bibnamefont
  {Belczynski}} \emph {et~al.},\ }\bibfield  {title} {\bibinfo {title} {{Binary
  neutron star formation and the origin of GW170817}},\ }\href@noop {}
  {\bibfield  {journal} {\bibinfo  {journal} {ArXiv e-print}\ } (\bibinfo
  {year} {2018})},\ \Eprint {https://arxiv.org/abs/1812.10065}
  {arXiv:1812.10065 [astro-ph.HE]} \BibitemShut {NoStop}%
\bibitem [{\citenamefont {Vigna-Gómez}\ \emph {et~al.}(2018)\citenamefont
  {Vigna-Gómez} \emph {et~al.}}]{Vigna-Gomez:2018dza}%
  \BibitemOpen
  \bibfield  {author} {\bibinfo {author} {\bibfnamefont {A.}~\bibnamefont
  {Vigna-Gómez}} \emph {et~al.},\ }\bibfield  {title} {\bibinfo {title} {{On
  the formation history of Galactic double neutron stars}},\ }\href
  {https://doi.org/10.1093/mnras/sty2463} {\bibfield  {journal} {\bibinfo
  {journal} {Mon.\ Not.\ Roy.\ Astron.\ Soc.}\ }\textbf {\bibinfo {volume}
  {481}},\ \bibinfo {pages} {4009} (\bibinfo {year} {2018})},\ \Eprint
  {https://arxiv.org/abs/1805.07974} {arXiv:1805.07974 [astro-ph.SR]}
  \BibitemShut {NoStop}%
\bibitem [{\citenamefont {Giacobbo}\  and\ \citenamefont
  {Mapelli}(2018)}]{Giacobbo:2018etu}%
  \BibitemOpen
  \bibfield  {author} {\bibinfo {author} {\bibfnamefont {N.}~\bibnamefont
  {Giacobbo}} and\ \bibinfo {author} {\bibfnamefont {M.}~\bibnamefont
  {Mapelli}},\ }\bibfield  {title} {\bibinfo {title} {{The progenitors of
  compact-object binaries: impact of metallicity, common envelope and natal
  kicks}},\ }\href {https://doi.org/10.1093/mnras/sty1999} {\bibfield
  {journal} {\bibinfo  {journal} {Mon.\ Not.\ Roy.\ Astron.\ Soc.}\ }\textbf
  {\bibinfo {volume} {480}},\ \bibinfo {pages} {2011} (\bibinfo {year}
  {2018})},\ \Eprint {https://arxiv.org/abs/1806.00001} {arXiv:1806.00001
  [astro-ph.HE]} \BibitemShut {NoStop}%
\bibitem [{\citenamefont {Mapelli}\  and\ \citenamefont
  {Giacobbo}(2018)}]{Mapelli:2018wys}%
  \BibitemOpen
  \bibfield  {author} {\bibinfo {author} {\bibfnamefont {M.}~\bibnamefont
  {Mapelli}} and\ \bibinfo {author} {\bibfnamefont {N.}~\bibnamefont
  {Giacobbo}},\ }\bibfield  {title} {\bibinfo {title} {{The cosmic merger rate
  of neutron stars and black holes}},\ }\href
  {https://doi.org/10.1093/mnras/sty1613} {\bibfield  {journal} {\bibinfo
  {journal} {Mon.\ Not.\ Roy.\ Astron.\ Soc.}\ }\textbf {\bibinfo {volume}
  {479}},\ \bibinfo {pages} {4391} (\bibinfo {year} {2018})},\ \Eprint
  {https://arxiv.org/abs/1806.04866} {arXiv:1806.04866 [astro-ph.HE]}
  \BibitemShut {NoStop}%
\bibitem [{\citenamefont {Andrews}\  and\ \citenamefont
  {Mandel}(2019)}]{Andrews:2019vou}%
  \BibitemOpen
  \bibfield  {author} {\bibinfo {author} {\bibfnamefont {J.~J.}\ \bibnamefont
  {Andrews}} and\ \bibinfo {author} {\bibfnamefont {I.}~\bibnamefont
  {Mandel}},\ }\bibfield  {title} {\bibinfo {title} {{Double Neutron Star
  Populations and Formation Channels}},\ }\href
  {https://doi.org/10.3847/2041-8213/ab2ed1} {\bibfield  {journal} {\bibinfo
  {journal} {Astrophys. J.}\ }\textbf {\bibinfo {volume} {880}},\ \bibinfo
  {pages} {L8} (\bibinfo {year} {2019})},\ \Eprint
  {https://arxiv.org/abs/1904.12745} {arXiv:1904.12745 [astro-ph.HE]}
  \BibitemShut {NoStop}%
\bibitem [{\citenamefont {Lee}\ \emph {et~al.}(2010)\citenamefont {Lee},
  \citenamefont {Ramirez-Ruiz},\  and\ \citenamefont {van~de
  Ven}}]{Lee:2009ca}%
  \BibitemOpen
  \bibfield  {author} {\bibinfo {author} {\bibfnamefont {W.~H.}\ \bibnamefont
  {Lee}}, \bibinfo {author} {\bibfnamefont {E.}~\bibnamefont {Ramirez-Ruiz}},
  and\ \bibinfo {author} {\bibfnamefont {G.}~\bibnamefont {van~de Ven}},\
  }\bibfield  {title} {\bibinfo {title} {{Short gamma-ray bursts from
  dynamically-assembled compact binaries in globular clusters: pathways, rates,
  hydrodynamics and cosmological setting}},\ }\href
  {https://doi.org/10.1088/0004-637X/720/1/953} {\bibfield  {journal} {\bibinfo
   {journal} {Astrophys. J.}\ }\textbf {\bibinfo {volume} {720}},\ \bibinfo
  {pages} {953} (\bibinfo {year} {2010})},\ \Eprint
  {https://arxiv.org/abs/0909.2884} {arXiv:0909.2884 [astro-ph.HE]}
  \BibitemShut {NoStop}%
\bibitem [{\citenamefont {Ye}\ \emph {et~al.}(2020)\citenamefont {Ye},
  \citenamefont {Fong}, \citenamefont {Kremer}, \citenamefont {Rodriguez},
  \citenamefont {Chatterjee}, \citenamefont {Fragione},\  and\ \citenamefont
  {Rasio}}]{Ye:2019xvf}%
  \BibitemOpen
  \bibfield  {author} {\bibinfo {author} {\bibfnamefont {C.~S.}\ \bibnamefont
  {Ye}}, \bibinfo {author} {\bibfnamefont {W.-f.}\ \bibnamefont {Fong}},
  \bibinfo {author} {\bibfnamefont {K.}~\bibnamefont {Kremer}}, \bibinfo
  {author} {\bibfnamefont {C.~L.}\ \bibnamefont {Rodriguez}}, \bibinfo {author}
  {\bibfnamefont {S.}~\bibnamefont {Chatterjee}}, \bibinfo {author}
  {\bibfnamefont {G.}~\bibnamefont {Fragione}},  and\ \bibinfo {author}
  {\bibfnamefont {F.~A.}\ \bibnamefont {Rasio}},\ }\bibfield  {title} {\bibinfo
  {title} {{On the Rate of Neutron Star Binary Mergers from Globular
  Clusters}},\ }\href {https://doi.org/10.3847/2041-8213/ab5dc5} {\bibfield
  {journal} {\bibinfo  {journal} {Astrophys. J. Lett.}\ }\textbf {\bibinfo
  {volume} {888}},\ \bibinfo {pages} {L10} (\bibinfo {year} {2020})},\ \Eprint
  {https://arxiv.org/abs/1910.10740} {arXiv:1910.10740 [astro-ph.HE]}
  \BibitemShut {NoStop}%
\bibitem [{\citenamefont {Abbott}\ \emph {et~al.}(2017)\citenamefont {Abbott}
  \emph {et~al.}}]{TheLIGOScientific:2017qsa}%
  \BibitemOpen
  \bibfield  {author} {\bibinfo {author} {\bibfnamefont {B.~P.}\ \bibnamefont
  {Abbott}} \emph {et~al.} (\bibinfo {collaboration} {LIGO Scientific,
  Virgo}),\ }\bibfield  {title} {\bibinfo {title} {{GW170817: Observation of
  Gravitational Waves from a Binary Neutron Star Inspiral}},\ }\href
  {https://doi.org/10.1103/PhysRevLett.119.161101} {\bibfield  {journal}
  {\bibinfo  {journal} {Phys. Rev. Lett.}\ }\textbf {\bibinfo {volume} {119}},\
  \bibinfo {pages} {161101} (\bibinfo {year} {2017})},\ \Eprint
  {https://arxiv.org/abs/1710.05832} {arXiv:1710.05832 [gr-qc]} \BibitemShut
  {NoStop}%
\bibitem [{\citenamefont {Abbott}\ \emph {et~al.}(2020)\citenamefont {Abbott}
  \emph {et~al.}}]{Abbott:2020uma}%
  \BibitemOpen
  \bibfield  {author} {\bibinfo {author} {\bibfnamefont {B.}~\bibnamefont
  {Abbott}} \emph {et~al.} (\bibinfo {collaboration} {LIGO Scientific,
  Virgo}),\ }\bibfield  {title} {\bibinfo {title} {{GW190425: Observation of a
  Compact Binary Coalescence with Total Mass $\sim 3.4 M_{\odot}$}},\ }\href
  {https://doi.org/10.3847/2041-8213/ab75f5} {\bibfield  {journal} {\bibinfo
  {journal} {Astrophys. J. Lett.}\ }\textbf {\bibinfo {volume} {892}},\
  \bibinfo {pages} {L3} (\bibinfo {year} {2020})},\ \Eprint
  {https://arxiv.org/abs/2001.01761} {arXiv:2001.01761 [astro-ph.HE]}
  \BibitemShut {NoStop}%
\bibitem [{\citenamefont {Allen}(2005)}]{Allen:2004gu}%
  \BibitemOpen
  \bibfield  {author} {\bibinfo {author} {\bibfnamefont {B.}~\bibnamefont
  {Allen}},\ }\bibfield  {title} {\bibinfo {title} {{${\chi}^{2}$
  time-frequency discriminator for gravitational wave detection}},\ }\href
  {https://doi.org/10.1103/PhysRevD.71.062001} {\bibfield  {journal} {\bibinfo
  {journal} {Phys. Rev. D}\ }\textbf {\bibinfo {volume} {71}},\ \bibinfo
  {pages} {062001} (\bibinfo {year} {2005})},\ \Eprint
  {https://arxiv.org/abs/gr-qc/0405045} {arXiv:gr-qc/0405045} \BibitemShut
  {NoStop}%
\bibitem [{\citenamefont {Allen}\ \emph {et~al.}(2012)\citenamefont {Allen},
  \citenamefont {Anderson}, \citenamefont {Brady}, \citenamefont {Brown},\
  and\ \citenamefont {Creighton}}]{Allen:2005fk}%
  \BibitemOpen
  \bibfield  {author} {\bibinfo {author} {\bibfnamefont {B.}~\bibnamefont
  {Allen}}, \bibinfo {author} {\bibfnamefont {W.~G.}\ \bibnamefont {Anderson}},
  \bibinfo {author} {\bibfnamefont {P.~R.}\ \bibnamefont {Brady}}, \bibinfo
  {author} {\bibfnamefont {D.~A.}\ \bibnamefont {Brown}},  and\ \bibinfo
  {author} {\bibfnamefont {J.~D.~E.}\ \bibnamefont {Creighton}},\ }\bibfield
  {title} {\bibinfo {title} {{FINDCHIRP: An Algorithm for detection of
  gravitational waves from inspiraling compact binaries}},\ }\href
  {https://doi.org/10.1103/PhysRevD.85.122006} {\bibfield  {journal} {\bibinfo
  {journal} {Phys. Rev.}\ }\textbf {\bibinfo {volume} {D85}},\ \bibinfo {pages}
  {122006} (\bibinfo {year} {2012})},\ \Eprint
  {https://arxiv.org/abs/gr-qc/0509116} {arXiv:gr-qc/0509116 [gr-qc]}
  \BibitemShut {NoStop}%
\bibitem [{\citenamefont {Dal~Canton}\ \emph {et~al.}(2014)\citenamefont
  {Dal~Canton} \emph {et~al.}}]{Canton:2014ena}%
  \BibitemOpen
  \bibfield  {author} {\bibinfo {author} {\bibfnamefont {T.}~\bibnamefont
  {Dal~Canton}} \emph {et~al.},\ }\bibfield  {title} {\bibinfo {title}
  {{Implementing a search for aligned-spin neutron star-black hole systems with
  advanced ground based gravitational wave detectors}},\ }\href
  {https://doi.org/10.1103/PhysRevD.90.082004} {\bibfield  {journal} {\bibinfo
  {journal} {Phys. Rev. D}\ }\textbf {\bibinfo {volume} {90}},\ \bibinfo
  {pages} {082004} (\bibinfo {year} {2014})},\ \Eprint
  {https://arxiv.org/abs/1405.6731} {arXiv:1405.6731 [gr-qc]} \BibitemShut
  {NoStop}%
\bibitem [{\citenamefont {Usman}\ \emph {et~al.}(2016)\citenamefont {Usman}
  \emph {et~al.}}]{Usman:2015kfa}%
  \BibitemOpen
  \bibfield  {author} {\bibinfo {author} {\bibfnamefont {S.~A.}\ \bibnamefont
  {Usman}} \emph {et~al.},\ }\bibfield  {title} {\bibinfo {title} {{The PyCBC
  search for gravitational waves from compact binary coalescence}},\ }\href
  {https://doi.org/10.1088/0264-9381/33/21/215004} {\bibfield  {journal}
  {\bibinfo  {journal} {Class. Quant. Grav.}\ }\textbf {\bibinfo {volume}
  {33}},\ \bibinfo {pages} {215004} (\bibinfo {year} {2016})},\ \Eprint
  {https://arxiv.org/abs/1508.02357} {arXiv:1508.02357 [gr-qc]} \BibitemShut
  {NoStop}%
\bibitem [{\citenamefont {Nitz}\ \emph {et~al.}(2017)\citenamefont {Nitz},
  \citenamefont {Dent}, \citenamefont {Dal~Canton}, \citenamefont {Fairhurst},\
   and\ \citenamefont {Brown}}]{Nitz:2017svb}%
  \BibitemOpen
  \bibfield  {author} {\bibinfo {author} {\bibfnamefont {A.~H.}\ \bibnamefont
  {Nitz}}, \bibinfo {author} {\bibfnamefont {T.}~\bibnamefont {Dent}}, \bibinfo
  {author} {\bibfnamefont {T.}~\bibnamefont {Dal~Canton}}, \bibinfo {author}
  {\bibfnamefont {S.}~\bibnamefont {Fairhurst}},  and\ \bibinfo {author}
  {\bibfnamefont {D.~A.}\ \bibnamefont {Brown}},\ }\bibfield  {title} {\bibinfo
  {title} {{Detecting binary compact-object mergers with gravitational waves:
  Understanding and Improving the sensitivity of the PyCBC search}},\ }\href
  {https://doi.org/10.3847/1538-4357/aa8f50} {\bibfield  {journal} {\bibinfo
  {journal} {Astrophys. J.}\ }\textbf {\bibinfo {volume} {849}},\ \bibinfo
  {pages} {118} (\bibinfo {year} {2017})},\ \Eprint
  {https://arxiv.org/abs/1705.01513} {arXiv:1705.01513 [gr-qc]} \BibitemShut
  {NoStop}%
\bibitem [{\citenamefont {Sachdev}\ \emph {et~al.}(2019)\citenamefont {Sachdev}
  \emph {et~al.}}]{Sachdev:2019vvd}%
  \BibitemOpen
  \bibfield  {author} {\bibinfo {author} {\bibfnamefont {S.}~\bibnamefont
  {Sachdev}} \emph {et~al.},\ }\bibfield  {title} {\bibinfo {title} {{The
  GstLAL Search Analysis Methods for Compact Binary Mergers in Advanced LIGO's
  Second and Advanced Virgo's First Observing Runs}},\ }\href@noop {}
  {\bibfield  {journal} {\bibinfo  {journal} {ArXiv e-print}\ } (\bibinfo
  {year} {2019})},\ \Eprint {https://arxiv.org/abs/1901.08580}
  {arXiv:1901.08580 [gr-qc]} \BibitemShut {NoStop}%
\bibitem [{\citenamefont {Cannon}\ \emph {et~al.}(2020)\citenamefont {Cannon}
  \emph {et~al.}}]{Cannon:2020qnf}%
  \BibitemOpen
  \bibfield  {author} {\bibinfo {author} {\bibfnamefont {K.}~\bibnamefont
  {Cannon}} \emph {et~al.},\ }\bibfield  {title} {\bibinfo {title} {{GstLAL: A
  software framework for gravitational wave discovery}},\ }\href@noop {}
  {\bibfield  {journal} {\bibinfo  {journal} {ArXiv e-print}\ } (\bibinfo
  {year} {2020})},\ \Eprint {https://arxiv.org/abs/2010.05082}
  {arXiv:2010.05082 [astro-ph.IM]} \BibitemShut {NoStop}%
\bibitem [{\citenamefont {Davies}\ \emph {et~al.}(2020)\citenamefont {Davies},
  \citenamefont {Dent}, \citenamefont {T\'apai}, \citenamefont {Harry},
  \citenamefont {McIsaac},\  and\ \citenamefont {Nitz}}]{Davies:2020tsx}%
  \BibitemOpen
  \bibfield  {author} {\bibinfo {author} {\bibfnamefont {G.~S.}\ \bibnamefont
  {Davies}}, \bibinfo {author} {\bibfnamefont {T.}~\bibnamefont {Dent}},
  \bibinfo {author} {\bibfnamefont {M.}~\bibnamefont {T\'apai}}, \bibinfo
  {author} {\bibfnamefont {I.}~\bibnamefont {Harry}}, \bibinfo {author}
  {\bibfnamefont {C.}~\bibnamefont {McIsaac}},  and\ \bibinfo {author}
  {\bibfnamefont {A.~H.}\ \bibnamefont {Nitz}},\ }\bibfield  {title} {\bibinfo
  {title} {{Extending the PyCBC search for gravitational waves from compact
  binary mergers to a global network}},\ }\href
  {https://doi.org/10.1103/PhysRevD.102.022004} {\bibfield  {journal} {\bibinfo
   {journal} {Phys. Rev. D}\ }\textbf {\bibinfo {volume} {102}},\ \bibinfo
  {pages} {022004} (\bibinfo {year} {2020})},\ \Eprint
  {https://arxiv.org/abs/2002.08291} {arXiv:2002.08291 [astro-ph.HE]}
  \BibitemShut {NoStop}%
\bibitem [{\citenamefont {Dal~Canton}\ \emph {et~al.}(2020)\citenamefont
  {Dal~Canton}, \citenamefont {Nitz}, \citenamefont {Gadre}, \citenamefont
  {Davies}, \citenamefont {Villa-Ortega}, \citenamefont {Dent}, \citenamefont
  {Harry},\  and\ \citenamefont {Xiao}}]{DalCanton:2020vpm}%
  \BibitemOpen
  \bibfield  {author} {\bibinfo {author} {\bibfnamefont {T.}~\bibnamefont
  {Dal~Canton}}, \bibinfo {author} {\bibfnamefont {A.~H.}\ \bibnamefont
  {Nitz}}, \bibinfo {author} {\bibfnamefont {B.}~\bibnamefont {Gadre}},
  \bibinfo {author} {\bibfnamefont {G.~S.}\ \bibnamefont {Davies}}, \bibinfo
  {author} {\bibfnamefont {V.}~\bibnamefont {Villa-Ortega}}, \bibinfo {author}
  {\bibfnamefont {T.}~\bibnamefont {Dent}}, \bibinfo {author} {\bibfnamefont
  {I.}~\bibnamefont {Harry}},  and\ \bibinfo {author} {\bibfnamefont
  {L.}~\bibnamefont {Xiao}},\ }\bibfield  {title} {\bibinfo {title} {{Realtime
  search for compact binary mergers in Advanced LIGO and Virgo's third
  observing run using PyCBC Live}},\ }\href@noop {} {\  (\bibinfo {year}
  {2020})},\ \Eprint {https://arxiv.org/abs/2008.07494} {arXiv:2008.07494
  [astro-ph.HE]} \BibitemShut {NoStop}%
\bibitem [{\citenamefont {Lenon}\ \emph {et~al.}(2020)\citenamefont {Lenon},
  \citenamefont {Nitz},\  and\ \citenamefont {Brown}}]{Lenon:2020oza}%
  \BibitemOpen
  \bibfield  {author} {\bibinfo {author} {\bibfnamefont {A.~K.}\ \bibnamefont
  {Lenon}}, \bibinfo {author} {\bibfnamefont {A.~H.}\ \bibnamefont {Nitz}},
  and\ \bibinfo {author} {\bibfnamefont {D.~A.}\ \bibnamefont {Brown}},\
  }\bibfield  {title} {\bibinfo {title} {{Measuring the eccentricity of
  GW170817 and GW190425}},\ }\href {https://doi.org/10.1093/mnras/staa2120}
  {\bibfield  {journal} {\bibinfo  {journal} {Mon. Not. Roy. Astron. Soc.}\
  }\textbf {\bibinfo {volume} {497}},\ \bibinfo {pages} {1966} (\bibinfo {year}
  {2020})},\ \Eprint {https://arxiv.org/abs/2005.14146} {arXiv:2005.14146
  [astro-ph.HE]} \BibitemShut {NoStop}%
\bibitem [{\citenamefont {Romero-Shaw}\ \emph {et~al.}(2019)\citenamefont
  {Romero-Shaw}, \citenamefont {Lasky},\  and\ \citenamefont
  {Thrane}}]{Romero-Shaw:2019itr}%
  \BibitemOpen
  \bibfield  {author} {\bibinfo {author} {\bibfnamefont {I.~M.}\ \bibnamefont
  {Romero-Shaw}}, \bibinfo {author} {\bibfnamefont {P.~D.}\ \bibnamefont
  {Lasky}},  and\ \bibinfo {author} {\bibfnamefont {E.}~\bibnamefont
  {Thrane}},\ }\bibfield  {title} {\bibinfo {title} {{Searching for
  Eccentricity: Signatures of Dynamical Formation in the First
  Gravitational-Wave Transient Catalogue of LIGO and Virgo}},\ }\href
  {https://doi.org/10.1093/mnras/stz2996} {\bibfield  {journal} {\bibinfo
  {journal} {Mon. Not. Roy. Astron. Soc.}\ }\textbf {\bibinfo {volume} {490}},\
  \bibinfo {pages} {5210} (\bibinfo {year} {2019})},\ \Eprint
  {https://arxiv.org/abs/1909.05466} {arXiv:1909.05466 [astro-ph.HE]}
  \BibitemShut {NoStop}%
\bibitem [{\citenamefont {Nitz}\ \emph {et~al.}(2019)\citenamefont {Nitz},
  \citenamefont {Lenon},\  and\ \citenamefont {Brown}}]{Nitz:2019spj}%
  \BibitemOpen
  \bibfield  {author} {\bibinfo {author} {\bibfnamefont {A.~H.}\ \bibnamefont
  {Nitz}}, \bibinfo {author} {\bibfnamefont {A.}~\bibnamefont {Lenon}},  and\
  \bibinfo {author} {\bibfnamefont {D.~A.}\ \bibnamefont {Brown}},\ }\bibfield
  {title} {\bibinfo {title} {{Search for Eccentric Binary Neutron Star Mergers
  in the first and second observing runs of Advanced LIGO}},\ }\href
  {https://doi.org/10.3847/1538-4357/ab6611} {\bibfield  {journal} {\bibinfo
  {journal} {Astrophys. J.}\ }\textbf {\bibinfo {volume} {890}},\ \bibinfo
  {pages} {1} (\bibinfo {year} {2019})},\ \Eprint
  {https://arxiv.org/abs/1912.05464} {arXiv:1912.05464 [astro-ph.HE]}
  \BibitemShut {NoStop}%
\bibitem [{\citenamefont {Abbott}\ \emph {et~al.}(2016)\citenamefont {Abbott}
  \emph {et~al.}}]{Aasi:2013wya}%
  \BibitemOpen
  \bibfield  {author} {\bibinfo {author} {\bibfnamefont {B.~P.}\ \bibnamefont
  {Abbott}} \emph {et~al.} (\bibinfo {collaboration} {LIGO Scientific
  Collaboration, Virgo Collaboration}),\ }\bibfield  {title} {\bibinfo {title}
  {{Prospects for Observing and Localizing Gravitational-Wave Transients with
  Advanced LIGO and Advanced Virgo}},\ }\href
  {https://doi.org/10.1007/lrr-2016-1} {\bibfield  {journal} {\bibinfo
  {journal} {Living Rev. Relat.}\ }\textbf {\bibinfo {volume} {19}},\ \bibinfo
  {pages} {1} (\bibinfo {year} {2016})},\ \Eprint
  {https://arxiv.org/abs/1304.0670} {arXiv:1304.0670 [gr-qc]} \BibitemShut
  {NoStop}%
\bibitem [{\citenamefont {Brown}\  and\ \citenamefont
  {Zimmerman}(2010)}]{Brown:2009ng}%
  \BibitemOpen
  \bibfield  {author} {\bibinfo {author} {\bibfnamefont {D.~A.}\ \bibnamefont
  {Brown}} and\ \bibinfo {author} {\bibfnamefont {P.~J.}\ \bibnamefont
  {Zimmerman}},\ }\bibfield  {title} {\bibinfo {title} {{The Effect of
  Eccentricity on Searches for Gravitational-Waves from Coalescing Compact
  Binaries in Ground-based Detectors}},\ }\href
  {https://doi.org/10.1103/PhysRevD.81.024007} {\bibfield  {journal} {\bibinfo
  {journal} {Phys. Rev.}\ }\textbf {\bibinfo {volume} {D81}},\ \bibinfo {pages}
  {024007} (\bibinfo {year} {2010})},\ \Eprint
  {https://arxiv.org/abs/0909.0066} {arXiv:0909.0066 [gr-qc]} \BibitemShut
  {NoStop}%
\bibitem [{\citenamefont {Huerta}\  and\ \citenamefont
  {Brown}(2013)}]{Huerta:2013qb}%
  \BibitemOpen
  \bibfield  {author} {\bibinfo {author} {\bibfnamefont {E.~A.}\ \bibnamefont
  {Huerta}} and\ \bibinfo {author} {\bibfnamefont {D.~A.}\ \bibnamefont
  {Brown}},\ }\bibfield  {title} {\bibinfo {title} {{Effect of eccentricity on
  binary neutron star searches in Advanced LIGO}},\ }\href
  {https://doi.org/10.1103/PhysRevD.87.127501} {\bibfield  {journal} {\bibinfo
  {journal} {Phys. Rev.}\ }\textbf {\bibinfo {volume} {D87}},\ \bibinfo {pages}
  {127501} (\bibinfo {year} {2013})},\ \Eprint
  {https://arxiv.org/abs/1301.1895} {arXiv:1301.1895 [gr-qc]} \BibitemShut
  {NoStop}%
\bibitem [{\citenamefont {Harry}\ \emph {et~al.}(2009)\citenamefont {Harry},
  \citenamefont {Allen},\  and\ \citenamefont {Sathyaprakash}}]{Harry:2009ea}%
  \BibitemOpen
  \bibfield  {author} {\bibinfo {author} {\bibfnamefont {I.~W.}\ \bibnamefont
  {Harry}}, \bibinfo {author} {\bibfnamefont {B.}~\bibnamefont {Allen}},  and\
  \bibinfo {author} {\bibfnamefont {B.~S.}\ \bibnamefont {Sathyaprakash}},\
  }\bibfield  {title} {\bibinfo {title} {{A Stochastic template placement
  algorithm for gravitational wave data analysis}},\ }\href
  {https://doi.org/10.1103/PhysRevD.80.104014} {\bibfield  {journal} {\bibinfo
  {journal} {Phys. Rev. D}\ }\textbf {\bibinfo {volume} {80}},\ \bibinfo
  {pages} {104014} (\bibinfo {year} {2009})},\ \Eprint
  {https://arxiv.org/abs/0908.2090} {arXiv:0908.2090 [gr-qc]} \BibitemShut
  {NoStop}%
\bibitem [{\citenamefont {Manca}\  and\ \citenamefont
  {Vallisneri}(2010)}]{Manca:2009xw}%
  \BibitemOpen
  \bibfield  {author} {\bibinfo {author} {\bibfnamefont {G.~M.}\ \bibnamefont
  {Manca}} and\ \bibinfo {author} {\bibfnamefont {M.}~\bibnamefont
  {Vallisneri}},\ }\bibfield  {title} {\bibinfo {title} {{Cover art: Issues in
  the metric-guided and metric-less placement of random and stochastic template
  banks}},\ }\href {https://doi.org/10.1103/PhysRevD.81.024004} {\bibfield
  {journal} {\bibinfo  {journal} {Phys. Rev. D}\ }\textbf {\bibinfo {volume}
  {81}},\ \bibinfo {pages} {024004} (\bibinfo {year} {2010})},\ \Eprint
  {https://arxiv.org/abs/0909.0563} {arXiv:0909.0563 [gr-qc]} \BibitemShut
  {NoStop}%
\bibitem [{\citenamefont {Reitze}\ \emph
  {et~al.}(2019{\natexlab{b}})\citenamefont {Reitze} \emph
  {et~al.}}]{Reitze:2019dyk}%
  \BibitemOpen
  \bibfield  {author} {\bibinfo {author} {\bibfnamefont {D.}~\bibnamefont
  {Reitze}} \emph {et~al.},\ }\bibfield  {title} {\bibinfo {title} {{The US
  Program in Ground-Based Gravitational Wave Science: Contribution from the
  LIGO Laboratory}},\ }\href@noop {} {\bibfield  {journal} {\bibinfo  {journal}
  {Bull. Am. Astron. Soc.}\ }\textbf {\bibinfo {volume} {51}},\ \bibinfo
  {pages} {141} (\bibinfo {year} {2019}{\natexlab{b}})},\ \Eprint
  {https://arxiv.org/abs/1903.04615} {arXiv:1903.04615 [astro-ph.IM]}
  \BibitemShut {NoStop}%
\bibitem [{\citenamefont {Kuns}\ \emph {et~al.}(2020)\citenamefont {Kuns},
  \citenamefont {Hall}, \citenamefont {Smith}, \citenamefont {Evans},
  \citenamefont {Fritschel}, \citenamefont {Wipf},\  and\ \citenamefont
  {Ballmer}}]{CE:NoiseCurves}%
  \BibitemOpen
  \bibfield  {author} {\bibinfo {author} {\bibfnamefont {K.}~\bibnamefont
  {Kuns}}, \bibinfo {author} {\bibfnamefont {E.}~\bibnamefont {Hall}}, \bibinfo
  {author} {\bibfnamefont {J.}~\bibnamefont {Smith}}, \bibinfo {author}
  {\bibfnamefont {M.}~\bibnamefont {Evans}}, \bibinfo {author} {\bibfnamefont
  {P.}~\bibnamefont {Fritschel}}, \bibinfo {author} {\bibfnamefont
  {C.}~\bibnamefont {Wipf}},  and\ \bibinfo {author} {\bibfnamefont
  {S.}~\bibnamefont {Ballmer}},\ }\href
  {https://dcc.cosmicexplorer.org/CE-T2000017/public} {\bibinfo {title} {Cosmic
  explorer sensitivity curves}} (\bibinfo {year} {2020})\BibitemShut {NoStop}%
\bibitem [{\citenamefont {Finn}(2001)}]{Finn:2000hj}%
  \BibitemOpen
  \bibfield  {author} {\bibinfo {author} {\bibfnamefont {L.~S.}\ \bibnamefont
  {Finn}},\ }\bibfield  {title} {\bibinfo {title} {{Aperture synthesis for
  gravitational wave data analysis: Deterministic sources}},\ }\href
  {https://doi.org/10.1103/PhysRevD.63.102001} {\bibfield  {journal} {\bibinfo
  {journal} {Phys. Rev. D}\ }\textbf {\bibinfo {volume} {63}},\ \bibinfo
  {pages} {102001} (\bibinfo {year} {2001})},\ \Eprint
  {https://arxiv.org/abs/gr-qc/0010033} {arXiv:gr-qc/0010033} \BibitemShut
  {NoStop}%
\bibitem [{\citenamefont {Maggiore}\ \emph {et~al.}(2020)\citenamefont
  {Maggiore} \emph {et~al.}}]{Maggiore:2019uih}%
  \BibitemOpen
  \bibfield  {author} {\bibinfo {author} {\bibfnamefont {M.}~\bibnamefont
  {Maggiore}} \emph {et~al.},\ }\bibfield  {title} {\bibinfo {title} {{Science
  Case for the Einstein Telescope}},\ }\href
  {https://doi.org/10.1088/1475-7516/2020/03/050} {\bibfield  {journal}
  {\bibinfo  {journal} {JCAP}\ }\textbf {\bibinfo {volume} {03}},\ \bibinfo
  {pages} {050}},\ \Eprint {https://arxiv.org/abs/1912.02622} {arXiv:1912.02622
  [astro-ph.CO]} \BibitemShut {NoStop}%
\bibitem [{\citenamefont {Owen}(1996)}]{Owen:1995tm}%
  \BibitemOpen
  \bibfield  {author} {\bibinfo {author} {\bibfnamefont {B.~J.}\ \bibnamefont
  {Owen}},\ }\bibfield  {title} {\bibinfo {title} {{Search templates for
  gravitational waves from inspiraling binaries: Choice of template spacing}},\
  }\href {https://doi.org/10.1103/PhysRevD.53.6749} {\bibfield  {journal}
  {\bibinfo  {journal} {Phys. Rev. D}\ }\textbf {\bibinfo {volume} {53}},\
  \bibinfo {pages} {6749} (\bibinfo {year} {1996})},\ \Eprint
  {https://arxiv.org/abs/gr-qc/9511032} {arXiv:gr-qc/9511032} \BibitemShut
  {NoStop}%
\bibitem [{\citenamefont {Owen}\  and\ \citenamefont
  {Sathyaprakash}(1999)}]{Owen:1998dk}%
  \BibitemOpen
  \bibfield  {author} {\bibinfo {author} {\bibfnamefont {B.~J.}\ \bibnamefont
  {Owen}} and\ \bibinfo {author} {\bibfnamefont {B.~S.}\ \bibnamefont
  {Sathyaprakash}},\ }\bibfield  {title} {\bibinfo {title} {{Matched filtering
  of gravitational waves from inspiraling compact binaries: Computational cost
  and template placement}},\ }\href
  {https://doi.org/10.1103/PhysRevD.60.022002} {\bibfield  {journal} {\bibinfo
  {journal} {Phys. Rev. D}\ }\textbf {\bibinfo {volume} {60}},\ \bibinfo
  {pages} {022002} (\bibinfo {year} {1999})},\ \Eprint
  {https://arxiv.org/abs/gr-qc/9808076} {arXiv:gr-qc/9808076} \BibitemShut
  {NoStop}%
\bibitem [{\citenamefont {Cokelaer}(2007)}]{Cokelaer:2007kx}%
  \BibitemOpen
  \bibfield  {author} {\bibinfo {author} {\bibfnamefont {T.}~\bibnamefont
  {Cokelaer}},\ }\bibfield  {title} {\bibinfo {title} {{Gravitational waves
  from inspiralling compact binaries: Hexagonal template placement and its
  efficiency in detecting physical signals}},\ }\href
  {https://doi.org/10.1103/PhysRevD.76.102004} {\bibfield  {journal} {\bibinfo
  {journal} {Phys. Rev. D}\ }\textbf {\bibinfo {volume} {76}},\ \bibinfo
  {pages} {102004} (\bibinfo {year} {2007})},\ \Eprint
  {https://arxiv.org/abs/0706.4437} {arXiv:0706.4437 [gr-qc]} \BibitemShut
  {NoStop}%
\bibitem [{\citenamefont {Brown}\ \emph {et~al.}(2012)\citenamefont {Brown},
  \citenamefont {Harry}, \citenamefont {Lundgren},\  and\ \citenamefont
  {Nitz}}]{Brown:2012qf}%
  \BibitemOpen
  \bibfield  {author} {\bibinfo {author} {\bibfnamefont {D.~A.}\ \bibnamefont
  {Brown}}, \bibinfo {author} {\bibfnamefont {I.}~\bibnamefont {Harry}},
  \bibinfo {author} {\bibfnamefont {A.}~\bibnamefont {Lundgren}},  and\
  \bibinfo {author} {\bibfnamefont {A.~H.}\ \bibnamefont {Nitz}},\ }\bibfield
  {title} {\bibinfo {title} {{Detecting binary neutron star systems with spin
  in advanced gravitational-wave detectors}},\ }\href
  {https://doi.org/10.1103/PhysRevD.86.084017} {\bibfield  {journal} {\bibinfo
  {journal} {Phys. Rev. D}\ }\textbf {\bibinfo {volume} {86}},\ \bibinfo
  {pages} {084017} (\bibinfo {year} {2012})},\ \Eprint
  {https://arxiv.org/abs/1207.6406} {arXiv:1207.6406 [gr-qc]} \BibitemShut
  {NoStop}%
\bibitem [{\citenamefont {Apostolatos}(1995)}]{Apostolatos:1995pj}%
  \BibitemOpen
  \bibfield  {author} {\bibinfo {author} {\bibfnamefont {T.~A.}\ \bibnamefont
  {Apostolatos}},\ }\bibfield  {title} {\bibinfo {title} {{Search templates for
  gravitational waves from precessing, inspiraling binaries}},\ }\href
  {https://doi.org/10.1103/PhysRevD.52.605} {\bibfield  {journal} {\bibinfo
  {journal} {Phys. Rev. D}\ }\textbf {\bibinfo {volume} {52}},\ \bibinfo
  {pages} {605} (\bibinfo {year} {1995})}\BibitemShut {NoStop}%
\bibitem [{\citenamefont {{LIGO Scientific Collaboration}}(2018)}]{lalsuite}%
  \BibitemOpen
  \bibfield  {author} {\bibinfo {author} {\bibnamefont {{LIGO Scientific
  Collaboration}}},\ }\href {https://doi.org/10.7935/GT1W-FZ16} {\bibinfo
  {title} {{LIGO} {A}lgorithm {L}ibrary - {LALS}uite}},\ \bibinfo
  {howpublished} {free software (GPL)} (\bibinfo {year} {2018})\BibitemShut
  {NoStop}%
\bibitem [{\citenamefont {Buonanno}\ \emph {et~al.}(2009)\citenamefont
  {Buonanno}, \citenamefont {Iyer}, \citenamefont {Ochsner}, \citenamefont
  {Pan},\  and\ \citenamefont {Sathyaprakash}}]{Buonanno:2009zt}%
  \BibitemOpen
  \bibfield  {author} {\bibinfo {author} {\bibfnamefont {A.}~\bibnamefont
  {Buonanno}}, \bibinfo {author} {\bibfnamefont {B.}~\bibnamefont {Iyer}},
  \bibinfo {author} {\bibfnamefont {E.}~\bibnamefont {Ochsner}}, \bibinfo
  {author} {\bibfnamefont {Y.}~\bibnamefont {Pan}},  and\ \bibinfo {author}
  {\bibfnamefont {B.~S.}\ \bibnamefont {Sathyaprakash}},\ }\bibfield  {title}
  {\bibinfo {title} {{Comparison of post-Newtonian templates for compact binary
  inspiral signals in gravitational-wave detectors}},\ }\href
  {https://doi.org/10.1103/PhysRevD.80.084043} {\bibfield  {journal} {\bibinfo
  {journal} {Phys. Rev.}\ }\textbf {\bibinfo {volume} {D80}},\ \bibinfo {pages}
  {084043} (\bibinfo {year} {2009})},\ \Eprint
  {https://arxiv.org/abs/0907.0700} {arXiv:0907.0700 [gr-qc]} \BibitemShut
  {NoStop}%
\bibitem [{\citenamefont {Bohé}\ \emph {et~al.}(2013)\citenamefont {Bohé},
  \citenamefont {Marsat},\  and\ \citenamefont {Blanchet}}]{Bohe:2013cla}%
  \BibitemOpen
  \bibfield  {author} {\bibinfo {author} {\bibfnamefont {A.}~\bibnamefont
  {Bohé}}, \bibinfo {author} {\bibfnamefont {S.}~\bibnamefont {Marsat}},  and\
  \bibinfo {author} {\bibfnamefont {L.}~\bibnamefont {Blanchet}},\ }\bibfield
  {title} {\bibinfo {title} {{Next-to-next-to-leading order spin–orbit
  effects in the gravitational wave flux and orbital phasing of compact
  binaries}},\ }\href {https://doi.org/10.1088/0264-9381/30/13/135009}
  {\bibfield  {journal} {\bibinfo  {journal} {Class. Quant. Grav.}\ }\textbf
  {\bibinfo {volume} {30}},\ \bibinfo {pages} {135009} (\bibinfo {year}
  {2013})},\ \Eprint {https://arxiv.org/abs/1303.7412} {arXiv:1303.7412
  [gr-qc]} \BibitemShut {NoStop}%
\bibitem [{\citenamefont {Mikoczi}\ \emph {et~al.}(2005)\citenamefont
  {Mikoczi}, \citenamefont {Vasuth},\  and\ \citenamefont
  {Gergely}}]{Mikoczi:2005dn}%
  \BibitemOpen
  \bibfield  {author} {\bibinfo {author} {\bibfnamefont {B.}~\bibnamefont
  {Mikoczi}}, \bibinfo {author} {\bibfnamefont {M.}~\bibnamefont {Vasuth}},
  and\ \bibinfo {author} {\bibfnamefont {L.~A.}\ \bibnamefont {Gergely}},\
  }\bibfield  {title} {\bibinfo {title} {{Self-interaction spin effects in
  inspiralling compact binaries}},\ }\href
  {https://doi.org/10.1103/PhysRevD.71.124043} {\bibfield  {journal} {\bibinfo
  {journal} {Phys. Rev.}\ }\textbf {\bibinfo {volume} {D71}},\ \bibinfo {pages}
  {124043} (\bibinfo {year} {2005})},\ \Eprint
  {https://arxiv.org/abs/astro-ph/0504538} {arXiv:astro-ph/0504538 [astro-ph]}
  \BibitemShut {NoStop}%
\bibitem [{\citenamefont {Arun}\ \emph {et~al.}(2009)\citenamefont {Arun},
  \citenamefont {Buonanno}, \citenamefont {Faye},\  and\ \citenamefont
  {Ochsner}}]{Arun:2008kb}%
  \BibitemOpen
  \bibfield  {author} {\bibinfo {author} {\bibfnamefont {K.~G.}\ \bibnamefont
  {Arun}}, \bibinfo {author} {\bibfnamefont {A.}~\bibnamefont {Buonanno}},
  \bibinfo {author} {\bibfnamefont {G.}~\bibnamefont {Faye}},  and\ \bibinfo
  {author} {\bibfnamefont {E.}~\bibnamefont {Ochsner}},\ }\bibfield  {title}
  {\bibinfo {title} {{Higher-order spin effects in the amplitude and phase of
  gravitational waveforms emitted by inspiraling compact binaries: Ready-to-use
  gravitational waveforms}},\ }\href
  {https://doi.org/10.1103/PhysRevD.79.104023, 10.1103/PhysRevD.84.049901}
  {\bibfield  {journal} {\bibinfo  {journal} {Phys. Rev.}\ }\textbf {\bibinfo
  {volume} {D79}},\ \bibinfo {pages} {104023} (\bibinfo {year} {2009})},\
  \bibinfo {note} {[Erratum: Phys. Rev.D84,049901(2011)]},\ \Eprint
  {https://arxiv.org/abs/0810.5336} {arXiv:0810.5336 [gr-qc]} \BibitemShut
  {NoStop}%
\bibitem [{\citenamefont {Moore}\ \emph {et~al.}(2016)\citenamefont {Moore},
  \citenamefont {Favata}, \citenamefont {Arun},\  and\ \citenamefont
  {Mishra}}]{Moore:2016qxz}%
  \BibitemOpen
  \bibfield  {author} {\bibinfo {author} {\bibfnamefont {B.}~\bibnamefont
  {Moore}}, \bibinfo {author} {\bibfnamefont {M.}~\bibnamefont {Favata}},
  \bibinfo {author} {\bibfnamefont {K.~G.}\ \bibnamefont {Arun}},  and\
  \bibinfo {author} {\bibfnamefont {C.~K.}\ \bibnamefont {Mishra}},\ }\bibfield
   {title} {\bibinfo {title} {{Gravitational-wave phasing for low-eccentricity
  inspiralling compact binaries to 3PN order}},\ }\href
  {https://doi.org/10.1103/PhysRevD.93.124061} {\bibfield  {journal} {\bibinfo
  {journal} {Phys. Rev.}\ }\textbf {\bibinfo {volume} {D93}},\ \bibinfo {pages}
  {124061} (\bibinfo {year} {2016})},\ \Eprint
  {https://arxiv.org/abs/1605.00304} {arXiv:1605.00304 [gr-qc]} \BibitemShut
  {NoStop}%
\bibitem [{\citenamefont {Nitz}\  and\ \citenamefont
  {Wang}(2021{\natexlab{a}})}]{Nitz:2020bdb}%
  \BibitemOpen
  \bibfield  {author} {\bibinfo {author} {\bibfnamefont {A.~H.}\ \bibnamefont
  {Nitz}} and\ \bibinfo {author} {\bibfnamefont {Y.-F.}\ \bibnamefont {Wang}},\
  }\bibfield  {title} {\bibinfo {title} {{Search for Gravitational Waves from
  High-Mass-Ratio Compact-Binary Mergers of Stellar Mass and Subsolar Mass
  Black Holes}},\ }\href {https://doi.org/10.1103/PhysRevLett.126.021103}
  {\bibfield  {journal} {\bibinfo  {journal} {Phys. Rev. Lett.}\ }\textbf
  {\bibinfo {volume} {126}},\ \bibinfo {pages} {021103} (\bibinfo {year}
  {2021}{\natexlab{a}})},\ \Eprint {https://arxiv.org/abs/2007.03583}
  {arXiv:2007.03583 [astro-ph.HE]} \BibitemShut {NoStop}%
\bibitem [{\citenamefont {Nitz}\  and\ \citenamefont
  {Wang}(2021{\natexlab{b}})}]{Nitz:2021mzz}%
  \BibitemOpen
  \bibfield  {author} {\bibinfo {author} {\bibfnamefont {A.~H.}\ \bibnamefont
  {Nitz}} and\ \bibinfo {author} {\bibfnamefont {Y.-F.}\ \bibnamefont {Wang}},\
  }\bibfield  {title} {\bibinfo {title} {{Search for gravitational waves from
  the coalescence of sub-solar mass and eccentric compact binaries}},\
  }\href@noop {} {\  (\bibinfo {year} {2021}{\natexlab{b}})},\ \Eprint
  {https://arxiv.org/abs/2102.00868} {arXiv:2102.00868 [astro-ph.HE]}
  \BibitemShut {NoStop}%
\bibitem [{\citenamefont {Baird}\ \emph {et~al.}(2013)\citenamefont {Baird},
  \citenamefont {Fairhurst}, \citenamefont {Hannam},\  and\ \citenamefont
  {Murphy}}]{Baird:2012cu}%
  \BibitemOpen
  \bibfield  {author} {\bibinfo {author} {\bibfnamefont {E.}~\bibnamefont
  {Baird}}, \bibinfo {author} {\bibfnamefont {S.}~\bibnamefont {Fairhurst}},
  \bibinfo {author} {\bibfnamefont {M.}~\bibnamefont {Hannam}},  and\ \bibinfo
  {author} {\bibfnamefont {P.}~\bibnamefont {Murphy}},\ }\bibfield  {title}
  {\bibinfo {title} {{Degeneracy between mass and spin in black-hole-binary
  waveforms}},\ }\href {https://doi.org/10.1103/PhysRevD.87.024035} {\bibfield
  {journal} {\bibinfo  {journal} {Phys.\ Rev.\ D}\ }\textbf {\bibinfo {volume}
  {87}},\ \bibinfo {pages} {024035} (\bibinfo {year} {2013})},\ \Eprint
  {https://arxiv.org/abs/1211.0546} {arXiv:1211.0546 [gr-qc]} \BibitemShut
  {NoStop}%
\bibitem [{\citenamefont {Martel}\  and\ \citenamefont
  {Poisson}(1999)}]{Martel:1999tm}%
  \BibitemOpen
  \bibfield  {author} {\bibinfo {author} {\bibfnamefont {K.}~\bibnamefont
  {Martel}} and\ \bibinfo {author} {\bibfnamefont {E.}~\bibnamefont
  {Poisson}},\ }\bibfield  {title} {\bibinfo {title} {{Gravitational waves from
  eccentric compact binaries: Reduction in signal-to-noise ratio due to
  nonoptimal signal processing}},\ }\href
  {https://doi.org/10.1103/PhysRevD.60.124008} {\bibfield  {journal} {\bibinfo
  {journal} {Phys. Rev. D}\ }\textbf {\bibinfo {volume} {60}},\ \bibinfo
  {pages} {124008} (\bibinfo {year} {1999})},\ \Eprint
  {https://arxiv.org/abs/gr-qc/9907006} {arXiv:gr-qc/9907006} \BibitemShut
  {NoStop}%
\bibitem [{\citenamefont {Tanay}\ \emph {et~al.}(2016)\citenamefont {Tanay},
  \citenamefont {Haney},\  and\ \citenamefont {Gopakumar}}]{Tanay:2016zog}%
  \BibitemOpen
  \bibfield  {author} {\bibinfo {author} {\bibfnamefont {S.}~\bibnamefont
  {Tanay}}, \bibinfo {author} {\bibfnamefont {M.}~\bibnamefont {Haney}},  and\
  \bibinfo {author} {\bibfnamefont {A.}~\bibnamefont {Gopakumar}},\ }\bibfield
  {title} {\bibinfo {title} {{Frequency and time domain inspiral templates for
  comparable mass compact binaries in eccentric orbits}},\ }\href
  {https://doi.org/10.1103/PhysRevD.93.064031} {\bibfield  {journal} {\bibinfo
  {journal} {Phys. Rev. D}\ }\textbf {\bibinfo {volume} {93}},\ \bibinfo
  {pages} {064031} (\bibinfo {year} {2016})},\ \Eprint
  {https://arxiv.org/abs/1602.03081} {arXiv:1602.03081 [gr-qc]} \BibitemShut
  {NoStop}%
\bibitem [{\citenamefont {Huerta}\ \emph {et~al.}(2017)\citenamefont {Huerta}
  \emph {et~al.}}]{Huerta:2016rwp}%
  \BibitemOpen
  \bibfield  {author} {\bibinfo {author} {\bibfnamefont {E.}~\bibnamefont
  {Huerta}} \emph {et~al.},\ }\bibfield  {title} {\bibinfo {title} {{Complete
  waveform model for compact binaries on eccentric orbits}},\ }\href
  {https://doi.org/10.1103/PhysRevD.95.024038} {\bibfield  {journal} {\bibinfo
  {journal} {Phys. Rev. D}\ }\textbf {\bibinfo {volume} {95}},\ \bibinfo
  {pages} {024038} (\bibinfo {year} {2017})},\ \Eprint
  {https://arxiv.org/abs/1609.05933} {arXiv:1609.05933 [gr-qc]} \BibitemShut
  {NoStop}%
\bibitem [{\citenamefont {Cao}\  and\ \citenamefont {Han}(2017)}]{Cao:2017ndf}%
  \BibitemOpen
  \bibfield  {author} {\bibinfo {author} {\bibfnamefont {Z.}~\bibnamefont
  {Cao}} and\ \bibinfo {author} {\bibfnamefont {W.-B.}\ \bibnamefont {Han}},\
  }\bibfield  {title} {\bibinfo {title} {{Waveform model for an eccentric
  binary black hole based on the effective-one-body-numerical-relativity
  formalism}},\ }\href {https://doi.org/10.1103/PhysRevD.96.044028} {\bibfield
  {journal} {\bibinfo  {journal} {Phys. Rev. D}\ }\textbf {\bibinfo {volume}
  {96}},\ \bibinfo {pages} {044028} (\bibinfo {year} {2017})},\ \Eprint
  {https://arxiv.org/abs/1708.00166} {arXiv:1708.00166 [gr-qc]} \BibitemShut
  {NoStop}%
\bibitem [{\citenamefont {Huerta}\ \emph {et~al.}(2018)\citenamefont {Huerta}
  \emph {et~al.}}]{Huerta:2017kez}%
  \BibitemOpen
  \bibfield  {author} {\bibinfo {author} {\bibfnamefont {E.~A.}\ \bibnamefont
  {Huerta}} \emph {et~al.},\ }\bibfield  {title} {\bibinfo {title} {{Eccentric,
  nonspinning, inspiral, Gaussian-process merger approximant for the detection
  and characterization of eccentric binary black hole mergers}},\ }\href
  {https://doi.org/10.1103/PhysRevD.97.024031} {\bibfield  {journal} {\bibinfo
  {journal} {Phys. Rev. D}\ }\textbf {\bibinfo {volume} {97}},\ \bibinfo
  {pages} {024031} (\bibinfo {year} {2018})},\ \Eprint
  {https://arxiv.org/abs/1711.06276} {arXiv:1711.06276 [gr-qc]} \BibitemShut
  {NoStop}%
\bibitem [{\citenamefont {Hinder}\ \emph {et~al.}(2018)\citenamefont {Hinder},
  \citenamefont {Kidder},\  and\ \citenamefont {Pfeiffer}}]{Hinder:2017sxy}%
  \BibitemOpen
  \bibfield  {author} {\bibinfo {author} {\bibfnamefont {I.}~\bibnamefont
  {Hinder}}, \bibinfo {author} {\bibfnamefont {L.~E.}\ \bibnamefont {Kidder}},
  and\ \bibinfo {author} {\bibfnamefont {H.~P.}\ \bibnamefont {Pfeiffer}},\
  }\bibfield  {title} {\bibinfo {title} {{Eccentric binary black hole
  inspiral-merger-ringdown gravitational waveform model from numerical
  relativity and post-Newtonian theory}},\ }\href
  {https://doi.org/10.1103/PhysRevD.98.044015} {\bibfield  {journal} {\bibinfo
  {journal} {Phys. Rev. D}\ }\textbf {\bibinfo {volume} {98}},\ \bibinfo
  {pages} {044015} (\bibinfo {year} {2018})},\ \Eprint
  {https://arxiv.org/abs/1709.02007} {arXiv:1709.02007 [gr-qc]} \BibitemShut
  {NoStop}%
\bibitem [{\citenamefont {Romero-Shaw}\ \emph {et~al.}(2020)\citenamefont
  {Romero-Shaw}, \citenamefont {Farrow}, \citenamefont {Thrane},\  and\ \citenamefont
  {Zhu}}]{Romero-Shaw:2020aaj}%
  \BibitemOpen
  \bibfield  {author} {\bibinfo {author} {\bibfnamefont {I.~M.}\ \bibnamefont
  {Romero-Shaw}}, \bibinfo {author} {\bibfnamefont {N.}\ \bibnamefont
  {Farrow}}, \bibinfo {author} {\bibfnamefont {E.}~\bibnamefont
  {Thrane}}, and\ \bibinfo {author} {\bibfnamefont {X.~J.}~\bibnamefont
  {Zhu}},\ }\bibfield  {title} {\bibinfo {title} {{On the origin of GW190425}},\ }\href
  {https://10.1093/mnrasl/slaa084} {\bibfield  {journal} {\bibinfo
  {journal} {Mon. Not. Roy. Astron. Soc.}\ }\textbf {\bibinfo {volume} {496}},\
  \bibinfo {pages} {L64} (\bibinfo {year} {2020})},\ \Eprint
  {https://arxiv.org/abs/2001.06492} {arXiv:2001.06492 [astro-ph.HE]}
  \BibitemShut {NoStop}%
\bibitem [{\citenamefont {Messick}\ \emph {et~al.}(2017)\citenamefont {Huerta}
  \emph {et~al.}}]{GSTLAL}%
  \BibitemOpen
  \bibfield  {author} {\bibinfo {author} {\bibfnamefont {C.}~\bibnamefont
  {Messick}}, \bibinfo {author} {\bibfnamefont {K.}\ \bibnamefont
  {Blackburn}}, \bibinfo {author} {\bibfnamefont {P.}~\bibnamefont
  {Brady}}, \bibinfo {author} {\bibfnamefont {P.}\ \bibnamefont
  {Brockill}}, \bibinfo {author} {\bibfnamefont {K.}~\bibnamefont
  {Cannon}}, and\ \bibinfo {author} {\bibfnamefont {R.}~\bibnamefont
  {Cariou}} \emph {et~al.},\ }\bibfield  {title} {\bibinfo {title} {{Analysis framework for the prompt discovery of compact binary mergers in gravitational-wave data}},\ }\href
  {https://doi.org/10.1103/PhysRevD.95.042001} {\bibfield  {journal} {\bibinfo
  {journal} {Phys. Rev. D}\ }\textbf {\bibinfo {volume} {95}},\ \bibinfo
  {pages} {042001} (\bibinfo {year} {2017})},\ \Eprint
  {https://arxiv.org/abs/1604.04324} {arXiv:1604.04324 [astro-ph.IM]} \BibitemShut
  {NoStop}%
 \bibitem [{\citenamefont {Adams}\ \emph {et~al.}(2016)\citenamefont {Adams}
  \emph {et~al.}}]{Adams:2015ulm}%
  \BibitemOpen
  \bibfield  {author} {\bibinfo {author} {\bibfnamefont {T.}\ \bibnamefont
  {Adams}} \emph {et~al.},\ }\bibfield  {title} {\bibinfo
  {title} {{Low-latency analysis pipeline for compact binary coalescences in the advanced gravitational wave detector era}},\ }\href {https://doi.org/10.1088/0264-9381/33/17/175012} {\bibfield
  {journal} {\bibinfo  {journal} {Class. Quant. Grav.}\ }\textbf {\bibinfo
  {volume} {33}},\ \bibinfo {pages} {175012} (\bibinfo {year} {2016})},\
  \Eprint {https://arxiv.org/abs/1512.02864} {arXiv:1512.02864 [gr-qc]}
  \BibitemShut {NoStop}%
\bibitem [{\citenamefont {Lower}\ \emph {et~al.}(2014)\citenamefont
  {Lower} \emph {et~al.}}]{Lower:2018seu}%
  \BibitemOpen
  \bibfield  {author} {\bibinfo {author} {\bibfnamefont {M.~E.}~\bibnamefont
  {Lower}}, \bibinfo {author} {\bibfnamefont {E.}\ \bibnamefont
  {Thrane}}, \bibinfo {author} {\bibfnamefont {P.~D.}~\bibnamefont
  {Lasky}}, and\ \bibinfo {author} {\bibfnamefont {R.}~\bibnamefont
  {Smith}},\ }\bibfield  {title} {\bibinfo {title}
  {{Measuring eccentricity in binary black hole inspirals with gravitational waves}},\ }\href
  {https://doi.org/10.1103/PhysRevD.98.083028} {\bibfield  {journal} {\bibinfo
  {journal} {Phys. Rev. D}\ }\textbf {\bibinfo {volume} {98}},\ \bibinfo
  {pages} {083028} (\bibinfo {year} {2018})},\ \Eprint
  {https://arxiv.org/abs/1806.05350} {arXiv:1806.05350 [astro-ph.HE]} \BibitemShut
  {NoStop}%
\bibitem [{\citenamefont {Lenon}\ \emph {et~al.}(2021)\citenamefont
  {Lower} \emph {et~al.}}]{cosmic-explorer-bns-data-release}%
  \BibitemOpen
  \bibfield  {author} {\bibinfo {author} {\bibfnamefont {A.~K.}\ \bibnamefont
  {Lenon}}, \bibinfo {author} {\bibfnamefont {A.~H.}\ \bibnamefont
  {Nitz}}, \bibinfo {author} and\ {\bibfnamefont {D.~A.}~\bibnamefont
  {Brown}},\ }\bibfield  {title} {\bibinfo {title}
  {{Data Release:Eccentric Binary Neutron Star Search Prospects for Cosmic Explorer}},} {\bibfield  {publisher} {\bibinfo {publisher} {GitHub}} (\bibinfo {year} {2021}),\ {\bibinfo {journal} {GitHub repository}},\ \bibinfo {howpublished} {\url{https://github.com/gwastro/cosmic-explorer-bns-eccentricity}}} \BibitemShut
  {NoStop}%
\bibitem [{\citenamefont {Huerta}(2014)}]{Huerta:2014eca}%
  \BibitemOpen
  \bibfield  {author} {\bibinfo {author} {\bibfnamefont {E.~A.}~\bibnamefont
  {Huerta}}, \bibinfo {author} {\bibfnamefont {P.}\ \bibnamefont
  {Kumar}}, \bibinfo {author} {\bibfnamefont {S.~T.}~\bibnamefont
  {McWilliams}}, \bibinfo {author} {\bibfnamefont {R.}~\bibnamefont
  {O'Shaughnessy}}, and\ \bibinfo {author} {\bibfnamefont {N.}~\bibnamefont
  {Yunes}},\ }\bibfield  {title} {\bibinfo {title} {{Accurate and efficient waveforms for compact binaries on eccentric orbits}},\ }\href
  {https://doi.org/10.1103/PhysRevD.90.084016} {\bibfield  {journal} {\bibinfo
  {journal} {Phys. Rev. D}\ }\textbf {\bibinfo {volume} {90}},\ \bibinfo
  {pages} {084016} (\bibinfo {year} {2014})},\ \Eprint
  {https://arxiv.org/abs/1408.3406} {arXiv:gr-qc/1408.3406} \BibitemShut
  {NoStop}%

\end{thebibliography}
\end{document}